\documentclass[hyper,11pt,paper]{article}
\usepackage{jheppub}
\usepackage[usenames,dvipsnames,table]{xcolor}
\usepackage{graphicx,amsmath,amssymb,multirow,array,bm,mathrsfs,bbold,bbm}
\usepackage{footmisc}
\usepackage{epsf,amsfonts,slashed,soul}
\usepackage{lipsum}
\usepackage{subfig}
\usepackage{pifont}
\usepackage{bbold}
\usepackage{tikz}
\usepackage{verbatim}
\usepackage{placeins}
\allowdisplaybreaks
 
  \newcommand{\R}[1]{{\textcolor{Green} {[R: #1]}}}

\newcommand{\be}{\begin{equation}}
\newcommand{\ee}{\end{equation}}
\newcommand{\ba}{\begin{eqnarray}}
\newcommand{\ea}{\end{eqnarray}}

\newcommand{\ie}{{\it{i.e.}}~}
\newcommand{\f}[2]{\frac{#1}{#2}}
\newcommand{\eq}[2]{\begin{equation} \label{eq:#1} #2 \end{equation}}
\newcommand{\abs}[1]{\lvert #1 \rvert}

\newcommand{\bc}[1]{\textcolor{red}{BF: #1}}

\def\ket#1{| #1 \rangle}

\newcommand{\cC}{{\mathcal{C}}}

\newcommand{\cO}{{\mathcal{O}}}
\newcommand{\cP}{{\mathcal{P}}}

\newcommand{\cS}{{\mathcal{S}}}

\title{How to Make Traversable Wormholes:\\ Eternal  AdS$_4$ Wormholes from Coupled CFT's}

\author[a]{Suzanne Bintanja,}
\author[a]{Ricardo Esp\'indola,}
\author[a,b]{Ben Freivogel}
\author[a]{and Dora Nikolakopoulou}
\affiliation[a]{Institute for Theoretical Physics, University of Amsterdam}
\affiliation[b]{GRAPPA, University of Amsterdam} 

\emailAdd{sbintanja@gmail.com }
\emailAdd{r.espindolaromero@uva.nl}
\emailAdd{benfreivogel@gmail.com}
\emailAdd{t.nikolakopoulou@uva.nl}

\abstract{
We construct an eternal traversable wormhole connecting two asymptotically $\text{AdS}_4$ regions. The wormhole is dual to the ground state of a system of two identical holographic CFT's coupled via a single low-dimension operator. The coupling between the two CFT's leads to negative null energy in the bulk, which  supports a static traversable wormhole. As the ground state of a simple Hamiltonian, it may be possible to make these wormholes in the lab or on a quantum computer.  
}

\begin{document}
\maketitle

\section{Introduction and Results}

Wormholes have been a puzzling topic for physicists for a century. 
Many efforts have been made to build traversable wormholes using different kinds of fields and techniques, most of which require either the insertion of exotic matter \cite{Morris:1988cz,Morris:1988tu,visser1989traversable,Visser:1989kg,Poisson:1995sv,Barcelo:1999hq,Visser:2003yf,blazquezsalcedo2020ellis} or higher derivative theories \cite{Bhawal:1992sz,Thibeault:2005ha,Arias:2010xg, Chernicoff:2020tvr} which lack UV completions \cite{camanho2016causality}. %

Recent work has shown how to build traversable wormholes in physically sensible theories. Gao, Jafferis, and Wall (GJW) \cite{GJW} showed how to make  asymptotically AdS black holes traversable for a short time by coupling the boundaries to each other. This approach has been extended in
 a number of other works since then \cite{Maldacena:2017axo,vanBreukelen:2017dul,deBoer:2018ibj,Almheiri:2018ijj,Bak:2018txn,Fu:2018oaq, Caceres:2018ehr,Hirano:2019ugo,deBoer:2019kyr,Fu:2019vco,Ben1,Freivogel:2019whb,May:2020tch,Fallows:2020ugr, Emparan-Marolf, Balushi-Marolf}. 
The first eternal traversable wormhole was constructed by Maldacena and Qi \cite{MQ} in asymptotically nearly-$\text{AdS}_2$ spacetime.
More recently, Maldacena, Milekhin and Popov (MMP) \cite{MMP} found a long-lived 4$D$ asymptotically flat traversable wormhole solution in the Standard Model (see also \cite{Maldacena:2020skw}). 

In this paper, we make use of the ingredients developed by GJW and MMP in order to construct an eternal traversable wormhole in asymptotically AdS$_4$ spacetime. Our motivation is twofold. First, by constructing wormholes in asymptotically AdS spacetime, we can use AdS/CFT to learn more about them. Second, our wormhole solution can be used to learn more about CFT's. To this end, we identify a family of Hamiltonians consisting of two copies of a CFT coupled by simple, local interactions whose ground state is dual to the traversable wormhole.

This last point is significant for constructing traversable wormholes in a lab or on a quantum computer. Some very interesting ideas on how to do this are described in \cite{Susskind:2017nto,Brown:2019hmk,Nezami:2021yaq}. Given access to a holographic CFT, one simply needs to implement the coupling and allow the system to cool to its ground state, which is dual to a traversable wormhole. 

Concretely, the  bulk theory we consider is described in Section \ref{sec:Fermions} and consists of Einstein-Maxwell theory with negative cosmological constant, a $U(1)$ gauge field and massless Dirac fermions coupled to the gauge field. A particular solution is the magnetically charged Reissner-Nordstr\"{o}m (RN) black hole. Due to the magnetic field, the charged fermions develop Landau levels. The lowest Landau level has exactly zero energy on the sphere, so we can think of them as effectively $2D$ fermionic degrees of freedom once we dimensionally reduce on the sphere.

The classical solution consists of two magnetically charged RN black holes connected through an Einstein-Rosen bridge which is non-traversable. The traversability of the wormhole is achieved by introducing a coupling between the two CFT's  (labelled $L, R$)
of the form
\be
\label{actionint1}
S_{\text{int}}=i \int d^3 x~h \left( \bar{\Psi}_-^R\Psi_+^L+ \bar{\Psi}_+^L\Psi_-^R\right)~.
\ee
Here $\Psi^R$ is the bulk field at the right boundary that is dual to the charged fermions in the right CFT, and $\Psi^L$ is defined analogously. Note that this is a local coupling involving  a single, low dimension operator in each CFT; this contrasts with the beautiful construction of Maldacena and Qi \cite{MQ} in the AdS$_2$ context, where a large number of operators must be coupled.

In Sections \ref{subsec:boundary conditions} and \ref{subsec:stress tensor}, we describe how this interaction has the effect of modifying the boundary conditions and the vacuum state. The stress tensor receives a  quantum correction of the form
\be\label{Tcasimirintro}
\langle T_{++}(x)\rangle = -\frac{1}{2\pi^3}\frac{q\lambda(h)}{ R^2} ~,
\ee
where $R$ is the sphere radius, $q$ is the charge of the black hole,  and $\lambda$ is given by (\ref{functionh}). For small coupling $h$, $\lambda(h) \approx h$, but our analysis remains valid for \emph{finite} $h$. 
A priori it is not clear whether a self-consistent solution exists in which the negative null energy supports a traversable wormhole. Since it is only the quantum correction that has a chance of making the wormhole traversable, the quantum effects have a large backreaction on the metric.

Typically, this would constitute an intractable problem: we cannot calculate the quantum state, and hence the stress tensor, until we know the geometry, but on the other hand we cannot solve the   Einstein equations to determine the geometry until we know the stress tensor.
In this case, 
we are able to self-consistently solve the system because the stress tensor takes a particularly simple form, depending locally on the metric (up to an overall factor).

In Section \ref{Sec:WHgeometry}, we discuss properties of both the linearized and non-linear solutions. The wormhole geometry has the following two regimes. The middle of the wormhole is nearly  AdS$_2\times \cS^2$. 
As we move away from the middle of the wormhole, the geometry smoothly interpolates to the near-extremal region of two RN black holes. Far away, the quantum contribution (\ref{Tcasimirintro}) becomes negligible and the geometry is that of two magnetically charged RN black holes (see Fig. \ref{Fig:wormhole}).

As a consequence of the boundary perturbation, the mass of the wormhole is slightly {\it decreased} by a term proportional to the  coupling
\be
M = M_{\text{ext}}+\Delta M~,\quad\text{with}\quad\Delta M \sim - \lambda^2(h)~.
\ee
An infalling observer will experience the geometry of a naked singularity as she approaches from infinity. However, there is no actual singularity as all of a sudden, deep in the throat region, the wormhole opens up and she comes out to the other side safely.

In Section \ref{sec:Hamiltonian}, we identify a simple Hamiltonian  whose ground state  is dual to the wormhole. The procedure is to begin with two identical holographic CFT's, each with a global $U(1)$ symmetry, so that they are dual to Einstein-Maxwell theory at low energies. We then turn on a chemical potential for each CFT separately, and turn on a coupling of the form $\mathbf{\bar{\Psi}}^R \mathbf{{\Psi}}^L$
where the $\mathbf{\Psi}$ operators are dual to a bulk massless charged fermion. 

Concretely, the Hamiltonian we analyze is
\be
H = H_L + H_R 
+ \mu (Q_L-Q_R) 
-\frac{ih}{\ell} \int  d\Omega_2 \left(\mathbf{\bar{\Psi}}_-^R\mathbf{\Psi}_+^L + \mathbf{\bar{\Psi}}_+^L \mathbf{\Psi}_-^R  \right)~,\\
\ee
This Hamiltonian is similar to the construction of Cottrell et al \cite{CFHL}. The authors showed that the Hamiltonian in their case has the thermofield double state as its ground state. That construction, however, did not have a semiclassical gravity dual.

We show  that the ground state of this theory is dual to our eternal traversable wormhole geometry for some range of the coupling $h$ and chemical potential $\mu$. We compare the wormhole to other  geometries with the same boundary conditions, which may dominate the ensemble. In particular, we consider two disconnected black holes and empty AdS. We compute  the ground state for different values of the parameters $h$ and $\mu$, and find that the wormhole is the ground state for $h>h_c$ and $\mu>\mu_c$, with the critical values given by
\be
h_c=\frac{\bar{r}^2}{G_Nq}\sqrt{\frac{2\pi}{3\cC}\left(1+\frac{2\bar{r}^2}{\ell^2}\right)} ~~ \text{and} ~~ \mu_c = \sqrt{\pi}m_p~,
\ee
with $m_p$ the Planck mass. Interestingly, as the non-local coupling vanishes, there is a triple point located at $h =0$, $\mu = \mu_c$ where the three phases meet.  For values $h <0$, the ground state is dominated by either empty AdS or the black hole phase.  

The challenge of building a traversable wormhole is to have enough negative energy to allow defocusing of null geodesics, allowing the sphere to contract and re-expand. Here we have added two ingredients so that the bulk dual remains semiclassical.
First, the chemical potential makes the decoupled system closer to being traversable, since the near-horizon geometry for an extremal black hole is $AdS_2 \times \mathcal{S}^2$, and thus, the size of the sphere is constant near the horizon. Therefore, a small amount of negative energy will allow the sphere to re-expand and render the wormhole traversable\footnote{We thank Daniel Jafferis for suggesting this approach.}. Second, by using bulk charged fermions in combination with a magnetically charged black hole, as was done in MMP \cite{MMP}, we enhance the negative energy due to the quantum effects. The key point is that a single 4d charged fermion acts like a large number $q$ of 2d light charged fields due to the large degeneracy of lowest Landau levels. 

{\bf Note:} We understand that overlapping results will appear in  \cite{souvik}. We thank S. Banerjee for discussions. Also, \cite{vanr} appeared very shortly before this work. There, asymptotically AdS$_4$ wormholes are also constructed, but with rather different ingredients. In addition, the solutions of \cite{vanr} have different symmetries than our solution: they preserve the full Poincare invariance in the boundary directions. It would be interesting to understand the relationship between the two constructions better. We thank M. van Raamsdonk for discussions.

\section{Massless fermions in AdS$_4$}\label{sec:Fermions}
We start this section by describing the particular theory of interest, as well as setting up the notation and conventions of spinors in curved space. Afterwards, we describe how the boundary conditions change once we couple the asymptotic boundaries. Finally, we compute the resulting stress tensor. 

\subsection{Dynamics}
The theory consists of Einstein-Maxwell gravity with matter described by the action
\be\label{action1}
S = \int d^4x \sqrt{g} \left( \frac{1}{16 \pi G_N}(R - 2 \Lambda) - \frac{1}{4g^2}F^2 + i \bar{\Psi} \slashed{D} \Psi \right)~.
\ee
In particular, we are considering a single massless Dirac fermion of charge one. 
In this section, we follow the approach and conventions of \cite{MMP}.

We consider $g$ to be small, so that loop corrections are suppressed.  A general class of spherically symmetric solutions with magnetic charge, denoted by the integer $q$, can be parametrized as follows
\be\label{ansatz1}
ds^2 = e^{2 \sigma(x,t)}(-dt^2 + dx^2) + R^2(x) ~d\Omega_2^2 ~, ~~~ A = \frac{q}{2} \cos \theta d \phi~.
\ee
Note that in this metric the range of $x$ is compact and fixing this range can be seen as a gauge choice. For now we use $x\in [0,\frac{\pi}{2}]$. To have a well-defined representation of the Clifford algebra at each point of the spacetime we introduce the vierbein
\be
e^1 = e^\sigma dt, ~~ e^2 = e^\sigma dx, ~~ e^3=R d\theta, ~~ e^4 = R \sin \theta d\phi~.
\ee
and by solving 
\be
de^a + \omega^{ab} \wedge e^b = 0, ~ ~~~ \omega^{ab} = - \omega^{ba}~,
\ee
we compute the spin connection components
\be
\omega^{12} = \sigma' dt + \dot{\sigma} dx, ~~ \omega^{32} = R' e^{- \sigma} d \theta, ~~ \omega^{42} = R' \sin \theta e^{- \sigma} d \phi, ~~ \omega^{43} = \cos \theta d \phi~.
\ee
Here a prime denotes a derivative with respect to $x$, while a dot denotes a derivative taken with respect to $t$.
We use the following basis for the gamma matrices in flat space
\be
\gamma^1 = i \sigma_x  \otimes 1, ~~ \gamma^2 = \sigma_y \otimes 1, ~~ \gamma^3 = \sigma_z \otimes \sigma_x, ~~ \gamma^4 = \sigma_z \otimes \sigma_y~.
\ee
In this basis the Dirac operator has the form
\be
\begin{split}
\slashed{D}  = & e^{-\sigma} \Big[i \sigma_x \Big(\partial_t + \frac{\dot{\sigma}}{2}\Big) + \sigma_y \Big(\partial_x + \frac{\sigma'}{2} + \frac{R'}{R} \Big) \Big] \otimes 1\\ 
& +  \frac{\sigma_z}{R} \otimes \Big[\sigma_y \frac{\partial_\phi - iA_\phi}{\sin \theta} + \sigma_x \Big(\partial_\theta + \frac{1}{2}\cot \theta \Big) \Big]~.
\end{split}
\ee
In the static case, the metric (\ref{ansatz1}) has two Killing vectors $\partial_t$ and $\partial_\phi$. Introducing the following ansatz will allow us to decompose in Fourier modes on the  sphere $\cS^2$,
\be\label{eq:bulkPsi}
\Psi = \frac{e^{-\frac{\sigma}{2}}}{R} \sum\limits_{m} \psi^m(t,x) \otimes \eta^m(\theta, \phi)~.
\ee
Here $\psi^m$ and $\eta^m$ are bi-spinors. In the rest of the paper we will suppress the indices on $\psi$. In this ansatz the Dirac equation is given by
\be
\begin{split}
\frac{e^{-\frac{3}{2}\sigma}}{R}\left(i\sigma_x\partial_t+\sigma_y\partial_x\right)\psi\otimes\eta&=-\lambda~,\\
\frac{e^{-\frac{\sigma}{2}}}{R^2}\sigma_z\psi\otimes\left(\sigma_y\frac{\partial_\phi-iA_\phi}{\sin(\theta)}+\sigma_x\left(\partial_\theta+\frac{1}{2}\cot(\theta)\right)\right)\eta&=\lambda~.\label{Diraceq}
\end{split}
\ee
 Restricting to the lowest Landau level decouples the equations and admits solutions of the form

\be
\label{ansatzps}
\psi_\pm= \sum\limits_k \alpha_k^{\pm} ~ e^{i k(t \mp x)}~,~~~ \eta_{\pm }^m = \Big(\sin \frac{\theta}{2} \Big)^{ j_\pm \pm m}  \Big(\cos \frac{\theta}{2} \Big)^{j_\pm \mp m } e^{im \phi}, ~~ j_{\pm} = \frac{1}{2}(-1 \mp q)~,
\ee
where $\psi_\pm$ are the components of $\psi$, and we choose $\sigma_z \eta_\pm = \pm \eta_\pm$ as the basis for $\eta$.
If we take $q>0$, the solution is 
\be\label{sol1}
\eta_{+}=0, ~~~ \eta_- =\sum_m \mathcal{C}^j_{m}\eta_-^m
, ~~~ -j \leq m \leq j~,
\ee
where we define the quantum number $j:=j_-$, so that in the lowest Landau level the degeneracy of the two-dimensional fields is $q$. The normalization constant is given by
\be\label{normalization1}
(\mathcal{C}_{m}^{j})^2 =\frac{1}{2 } \frac{\Gamma(2+2j)}{\Gamma(1+j-m) \Gamma(1+j+m)}~,
\ee
so that
\be
\int d^2 \Omega ~\bar{\eta}^{m_1} \eta^{m_2} = \delta_{m_1 m_2}~.
\ee

\subsection{Boundary conditions}
According to the AdS/CFT dictionary, a bulk Dirac spinor of mass $m$ is dual to a spin $1/2$ primary operator $\cO$ of conformal dimension 
\be
\Delta_{\pm} = \f32 \pm \sqrt{m^2 \ell^2}~,
\ee
where $\ell$ is the AdS radius \cite{HS, Henneaux}. The stability bound requires $m \geq 0$ \cite{AM}. When applying the correspondence, we should consider that the first order nature of the Dirac action goes hand in hand with the different dimensionality between the bulk and boundary spinors. The extrapolate dictionary instructs us to identify the two bulk chiral components with the same boundary field. In addition, when solving the Dirichlet boundary value problem, we should impose boundary conditions only on half of the spinor degrees of freedom. Our gamma matrix in the holographic radial direction satisfies $(\gamma^2)^2 = 1$ and $(\gamma^2)^\dagger = \gamma^2$. We can then decompose the bulk fermions onto the eigenspace of $\gamma^2$
\be
\Psi_{\pm} := \cP_{\pm} \Psi~,~~~~ \cP_{\pm} = \f12 \left(1 \pm \gamma^2 \right)~,
\ee
and similarly for the Dirac conjugate. The orthogonal projection operator satisfies the two conditions $\cP^2 = \cP$ and $\cP^\dagger=\cP$. More explicitly
\be
\begin{split}
\Psi_+&:=\frac{1}{2}\frac{e^{-\frac{\sigma}{2}}}{R}\begin{pmatrix}\psi_+-i\psi_- \\ i(\psi_+-i\psi_-)\end{pmatrix}\otimes\begin{pmatrix}\eta_+\\ \eta_-\end{pmatrix}~,\\
\Psi_-&:=\frac{1}{2}\frac{e^{-\frac{\sigma}{2}}}{R}\begin{pmatrix}\psi_++i\psi_- \\ -i(\psi_++i\psi_-)\end{pmatrix}\otimes\begin{pmatrix}\eta_+\\ \eta_-\end{pmatrix}~.
\end{split}
\ee
The variation of the Dirac part of the action (\ref{action1}) with respect to $\Psi_{\pm}$ after integration by parts becomes 
\begin{align}
\Delta S_D 
&  = {\rm bulk~terms} + i \int\limits_\partial d^3x~ \sqrt{\gamma} \left( \bar{\Psi}_- \delta \Psi_+ - \bar{\Psi}_+ \delta \Psi_- \right)~,
\end{align}
where $\gamma$ is the determinant of the induced metric at the boundary.
The bulk terms are proportional to the equations of motion. In order to have a well-defined boundary value problem, we should include a boundary term of the form
\be
S_\partial =  i \int \limits_\partial d^3x \sqrt{\gamma} \left( a_1 \bar{\Psi}_- \Psi_+ + a_2 \bar{\Psi}_+ \Psi_- \right)~. 
\ee
We can then either fix $ \Psi_+=0$ or $ \Psi_-=0$ (and thus $ \bar{\Psi}_+=0$ or $ \bar{\Psi}_-=0$) at the boundary depending on whether we set $(a_1=-1, a_2=0)$ or $(a_1=0, a_2=1)$ respectively in the total variation of the action $\delta S_D + \delta S_\partial$.

In the massless case, both modes $\Psi_\pm$ are normalizable. We can then identify the asymptotic values
\be\label{eq:extrapolate}
 \Psi_\pm^{0} := \lim\limits_{x \rightarrow \f\pi2 } R(x)^{-\f32} \Psi_\pm ~,
\ee
with the normalizable part of the dual operator $\cO$.
After reducing on the $\cS^2$ sphere, the effective $2D$ fermions obey reflective boundary conditions in both types of quantizations 
\be
\psi_+ = e^{i \alpha}~ \psi_-~,~~~ {\rm with} ~~~  \alpha =
\begin{cases}
    \f\pi2~,& \text{standard}\\
    \frac{3 \pi}{2}~,             & \text{alternate}
\end{cases}~,
\ee
which correspond to taking $\Psi_+^{0}=0$ or $\Psi_-^{0}=0$ respectively.
Intuitively, they would not allow the charge and energy to leak out at the boundary. In fact, by using the conservation equations it is easy to see that at the boundary
\be
\dot{E}  = T_{12}\Big\lvert_\partial =0~~~ {\rm and} ~~~ \dot{Q} = J_{2}\Big\lvert_\partial =0~,
\ee
where $J_2$ is the $x$ component of the $U(1)$ current 
\begin{equation}
    J_{2}=\psi^{\dagger}\sigma_{z}\psi \otimes\eta^{\dagger}\eta=\left(\psi_{-}^{\dagger}\psi_{+}-\psi_{+}^{\dagger}\psi_{-}\right)\otimes \eta^{\dagger}\eta~,
\end{equation}
and $T_{12}$ is the energy flux, which is given by the $tx$-component of the stress tensor
\be
T_{12}=\frac{i}{2R^2}\left(\psi_{+}^{\dagger}(\partial_x-\partial_t)\psi_{+}+\psi_{-}^{\dagger}(\partial_x+\partial_t)\psi_{-}-(\partial_x-\partial_t)\psi_{+}^{\dagger} \psi_{+}-(\partial_x+\partial_t)\psi_{-}^{\dagger}\psi_{-}\right)\otimes\eta^{\dagger}\eta~.
\ee
Now consider two decoupled and identical conformal theories with fermionic degrees of freedom. In principle, each one has its own bulk gravity dual. The boundary action then acquires the form\footnote{Note that since we consider two copies of the theory we now have $x\in [-\frac{\pi}{2},\frac{\pi}{2}]$, and we denote the left (right) boundary at $x=\mp\frac{\pi}{2}$ with $L$ $(R)$.}
\be
S_\partial = i \int\limits_\partial d^3 x \sqrt{\gamma} \left( a_1 \bar{\Psi}_-^R \Psi_+^R + a_2 \bar{\Psi}_+^R \Psi_-^R +  b_1 \bar{\Psi}_-^L \Psi_+^L + b_2 \bar{\Psi}_+^L \Psi_-^L \right)~.
\ee
There are various options depending on what type of boundary sources we would like to keep turned-on. The guiding principle we will follow is CPT invariance. CPT-related boundary conditions imply a vanishing $T_{++}$ component consistent with the fact that vacuum AdS$_2$ cannot support finite energy excitations \cite{Maldacena:1998uz}. For the purpose of this work, we choose the following CPT conjugate boundary conditions 

\be
\Psi_+^R = 0 ~ \xrightarrow{\text{CPT}} ~ \bar{\Psi}_-^L =0~,
\ee
which correspond to the coefficients $a_1=-1, \ b_2=1$, and $a_2=b_1=0$.
The vanishing energy can also be understood as due to the conformal anomaly contribution present in the mapping between the energy of a CFT on the strip to AdS$_2$ \cite{MQ}.

\subsection{Modified boundary conditions}\label{subsec:boundary conditions}

We are interested in the case where the two bulk geometries are two magnetically charged RN black holes. Intuitively, we can think on them as being connected through an Einstein-Rosen bridge. A priori, however, it is not obvious how to connect both bulk geometries through the horizon. Moreover, in order to render the wormhole traversable, we need to establish a connection between the two asymptotic boundaries. We achieve that by using a non-local coupling of the form\footnote{In general, the coupling constants can be complex. However, they must satisfy $h_1 = h_2^*, h_3 = h_4^*$, in order for (\ref{actionint2}) to be real.}
\be
\label{actionint2}
S_{\text{int}}= -i \int d^3x \sqrt{\gamma}  \left( h_1\bar{\Psi}_+^R\Psi_-^L+ h_2\bar{\Psi}_-^L\Psi_+^R + h_3\bar{\Psi}_-^R\Psi_+^L + h_4\bar{\Psi}_{+}^{L} \Psi_-^R \right)~.
\ee
 This term will provide us with the negative energy we need and will open up the wormhole. It is important to mention that if instead of fermions, we considered interacting scalar fields, similar to \cite{GJW}, the lowest Landau levels would have positive energy on the $\cS^2$ sphere, making the problem of finding a traversable geometry much harder. 
 
\begin{figure}[htbp]
\includegraphics[height=10cm]{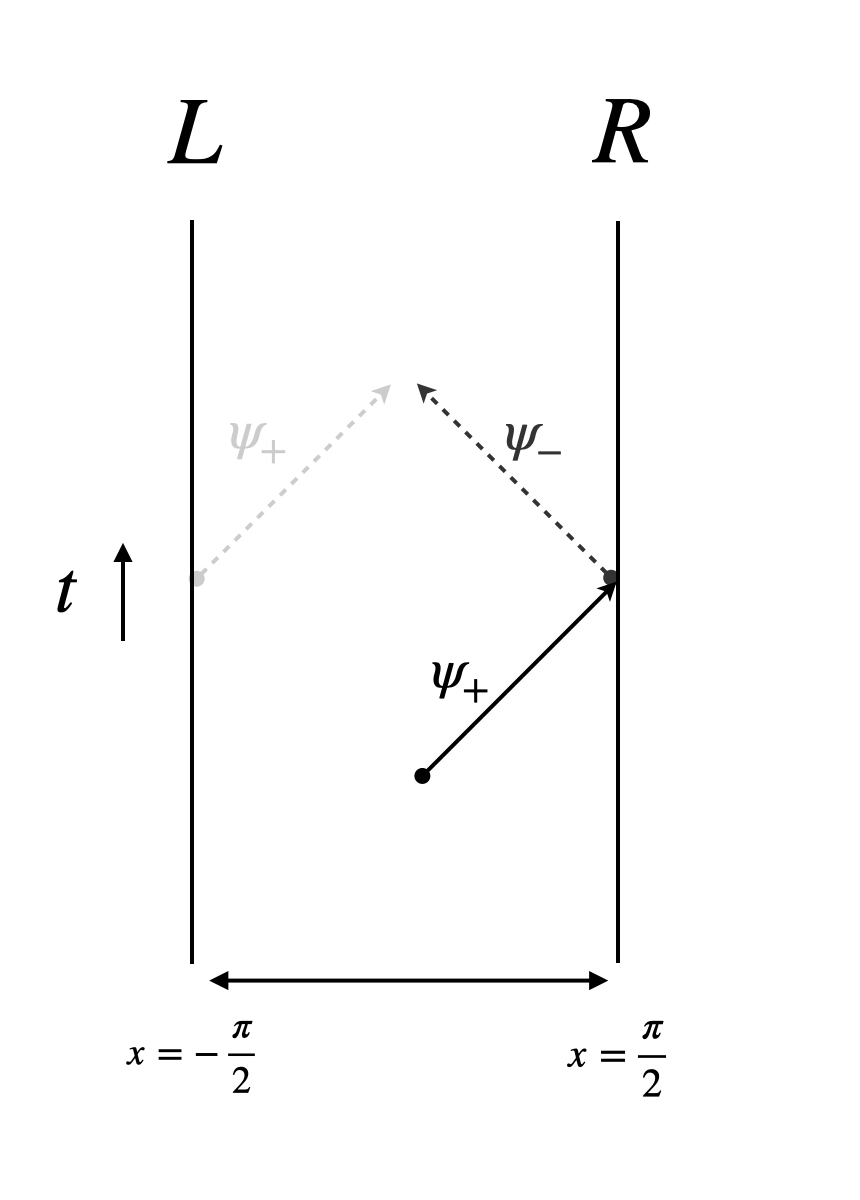}
\centering
\captionsetup{width=0.8\textwidth}
\caption{A right moving massless fermion, with amplitude $|\psi^{R}_{+}|=1$, traveling on the strip hits the right boundary. The probability of the resulting left mover is equal to $\frac{(h^2-1)^2}{(h^2+1)^2}$, and the right mover emerging from the left boundary has amplitude $\frac{4h^2}{(h^2+1)^2}$.}
\label{fig:fermions}
\end{figure}

We are looking for an eternal traversable wormhole, so we let the coupling constants be turned on for all times. For the purposes of this work, we focus on the case where the coupling constants are real and $h_1=h_2=0$\footnote{It would be interesting to understand other combinations of the non-local couplings.}. The boundary conditions turn out to be
\be
\label{modbdyconddelta}
\Psi_+^R+h\Psi_+^L=0,\quad\text{and}\quad\Psi_-^L+h\Psi_-^R=0~,
\ee
where $h_3=h_4=-     h$. Notice that the sources at the boundary are vanishing. In terms of the spinor components this implies the following boundary conditions
\begin{align}
\psi_+^R-i\psi_-^R+h\psi_+^L-ih\psi_-^L=0~,\quad\text{and}\quad
\psi_+^L+i\psi_-^L+h\psi_+^R+ih\psi_-^R=0~.\label{eq:modbdycond}
\end{align}
See Fig. \ref{fig:fermions} for an illustrations of the modified boundary conditions.

In order to obtain a solution to the equations of motion \eqref{Diraceq} with the boundary conditions \eqref{eq:modbdycond}, for the lowest Landau level, we use the following ansatz:
\be
\label{psiansatz}
\psi_+=\sum_k \frac{\alpha_k}{\sqrt{\pi}}e^{i\omega_k(t-x)}\quad\text{and}\quad\psi_-=\sum_k\frac{\beta_k}{\sqrt{\pi}}e^{i\omega_k(t+x)}~.
\ee
Filling in this ansatz in to the boundary conditions (\ref{eq:modbdycond}) leads to the following constraint equations\footnote{Note that the equations are invariant under $h\mapsto\frac{1}{h}$ and $\beta_k\mapsto-\beta_k$, so the theory exhibits S-duality.}
\be
\begin{split}
(i^{3\omega_k}+hi^{\omega_k})\alpha_k+(i^{\omega_k+3}+hi^{3\omega_k+3})\beta_k=&0~,\\
(i^{\omega_k}+hi^{3\omega_k})\alpha_k+(i^{3\omega_k+1}+hi^{\omega_k+1})\beta_k=&0~,
\end{split}
\ee
with solution
\be
\label{modfreq}
\omega_k=2k-\frac{i}{\pi}\log\left(\frac{-2h\pm i \lvert 1-h^2 \lvert}{1+h^2}\right),\quad\beta_k=(-1)^{k
+1}\alpha_k~,
\ee
where $k\in\mathbb{Z}$. The solution can be written in the following form
\be
\label{modfreqsmallh}
\omega_k=\frac{2k+1}{2}+(-1)^k {2 \over \pi} \lambda(h)~,
\ee
where $\lambda$ is a function of $h$ given by 
\be
\label{functionh}
\lambda(h)=\frac{1}{2}\arctan\left(\frac{2h}{\abs{1-h^2}}\right)~.
\ee

\subsection{Propagators and stress tensor}\label{subsec:stress tensor}
Using the solution \eqref{modfreqsmallh}, we write the fermionic fields as\footnote{In the remainder of this work, we will use light-cone coordinates defined by $x_{\pm}=t\pm x$, whenever they are more convenient.}

\be
\label{psisolution}
\psi_+=\sum_k\frac{1}{\sqrt{\pi}}\alpha_ke^{i\omega_k(t-x)}~,\quad\text{and}\quad\psi_-=\sum_k\frac{(-1)^{k+1}}{\sqrt{\pi}}\alpha_ke^{i\omega_k(t+x)}~.
\ee
The modes $\alpha_k$ obey the following anti-commutation relations
\be
\{\alpha_k,\alpha_j^\dagger\}=\delta_{k,j}~,\quad \{\alpha_k,\alpha_j\}=0\quad\text{and}\quad\ \{\alpha_k^\dagger,\alpha_j^\dagger\}=0~,
\ee
and the vacuum is defined as

\be
\label{modes}
\alpha_k\ket{0}=0\quad\forall\ k\in\mathbb{Z}_{<0},\quad\text{and}\quad\alpha_k^\dagger\ket{0}=0\quad\forall \ k\in\mathbb{Z}_{\ge0}~.
\ee
Using equations \eqref{psisolution}-\eqref{modes} we calculate the propagators. 
We present one of them here and the rest can be found in the Appendix \ref{App:Propagators}
\be
\label{prop}
\langle\psi_+^\dagger(x_-)\psi_+(x_-^\prime)\rangle=\frac{1}{\pi}\frac{e^{\frac{i}{2}(x_-^\prime-x_-)}}{1-e^{i(x_-^\prime-x_-)}}+\frac{e^{\frac{i}{2}(x_-^\prime-x_-)}}{1+e^{i(x_-^\prime-x_-)}}\frac{2i\lambda(h)}{\pi^2}(x_-^\prime-x_-)+\cdots~.
\ee

We proceed by stating the relevant components of the stress tensor. Since (\ref{ansatz1}) is spherically symmetric and does not depend on time, the only off-diagonal component of the stress tensor that could be nonzero is $T_{12}$. However, in Appendix \ref{sec:appT} we show explicitly that $T_{12}$ vanishes for our setup.  Therefore, we only need the diagonal components of the stress tensor, which are given by 
\be\label{stresstensorcomponents}
\begin{split}
&T_{11}=\frac{i}{2R^2}\left(\psi_+^\dagger\partial_t\psi_++\psi_-^\dagger\partial_t\psi_--\partial_t\psi_+^\dagger\psi_+-\partial_t\psi_-^\dagger\psi_-\right)\eta^\dagger\eta~,\\
&T_{22}=-\frac{i}{2R^2}\left(\psi_+^\dagger\partial_x\psi_+-\psi_-^\dagger\partial_x\psi_--\partial_x\psi_+^\dagger\psi_++\partial_x\psi_-^\dagger\psi_-\right)\eta^\dagger\eta~,\\
&T_{33}=-\frac{i e^{-2\sigma}}{2}\frac{R^\prime}{R}\psi^\dagger\sigma_z\psi\eta^\dagger\eta~,\\
&T_{44}=-\frac{i\sin(\theta)}{2}\frac{e^{-2\sigma}}{R}R^\prime\sin(\theta)e^{-\sigma}\psi^\dagger\sigma_z\psi\eta^\dagger\eta~.
\end{split}
\ee
 In order to compute the quantum contribution to the components of the stress tensor due to the non-local coupling, we apply the point-splitting formula
\be
\langle T_{\mu\nu}\rangle=\lim_{x^\prime\rightarrow x}\frac{i\eta_{ab}}{2}\left(e^a_{(\mu}\gamma^b\nabla^\prime_{\nu)}-\nabla_{(\mu}e_{\nu)}^a\gamma^b\right)\langle\bar{\Psi}(x)\Psi(x^\prime)\rangle~.
\ee
By using the propagators and after subtracting the vacuum contribution, we end up with the following finite result 
\be\label{Tcasimir}
\langle T^h_{\mu\nu}\rangle=-\frac{1}{2\pi^3}\frac{q\lambda(h)}{ R^2}\text{diag}\left(1,1,0,0\right)~,
\ee
where the factor $q$ comes from the fact that in the lowest Landau level the degeneracy of the two-dimensional fields is $q$. The range for the compact radial coordinate $(\Delta x = \pi)$ is present in the prefactor in the above expression. Picking a different gauge would result in a rescaling of the stress tensor. One can easily check that the stress tensor is conserved and traceless due to conformal symmetry. Details of the stress tensor calculation can be found in Appendix \ref{sec:appT}.

\section{Wormhole geometry}\label{Sec:WHgeometry}

We start this section by describing the two different regimes of the wormhole geometry, after which we analytically solve the (linearized) Einstein equations in both regimes. We continue by showing that the solutions in the two regimes can be consistently patched together through a coordinate transformation in the overlapping region of validity. We end the section by solving the full, nonlinear Einstein equations numerically. 
\subsection{Two regimes}
The next task is to solve the semi-classical Einstein equations to find a magnetically charged geometry sourced by (\ref{Tcasimir}) 
\be
\label{Einstein}
G_{\mu \nu} + \Lambda g_{\mu \nu} = 8 \pi G_N \langle T_{\mu \nu}\rangle~.
\ee
As we approach the AdS$_4$ boundaries located at $r \rightarrow \pm \infty$, the electromagnetic contribution of the stress tensor dominates over the Casimir energy. Then, far away from the wormhole throat the solution should look like Reissner-Nordstr\"{o}m AdS$_4$
\be\label{RNmetric}
d s^2=-f(r)d \tau^2+\frac{d r^2}{f(r)}+r^2 d \Omega_2^2~,
\ee
with emblackening factor
\be\label{fbh}
f(r)=1-\frac{2G_NM}{r}+\frac{r_e^2}{r^2}+\frac{r^2}{\ell^2}, ~~~ {\rm{and}} ~~~ r_e^2=\frac{\pi q^2G_N}{g^2}~.
\ee
Here $M$ denotes the mass of the black hole, $q$ is an integer and $r_e^2$ denotes the magnetic charge of the black hole. Close to extremality, the geometry developes an infinitely long throat. The value of the extremal radius has the form
\be
\label{extremalcond2}
\partial_rf(r) \Big\lvert_{r= \bar{r}} \overset{!}=0 ~~\Rightarrow ~~\bar{r}^2=\frac{\ell^2}{6}\left(-1+\sqrt{1+12\frac{r_e^2}{\ell^2}}\right)~.
\ee
Inverting this relation gives the charge of the black hole in terms of the extremal horizon radius and the AdS length 
\be
r_e^2=\bar{r}^2\left(1+3\frac{\bar{r}^2}{\ell^2}\right).
\ee
In the range of masses that we are interested in, the quartic polynomial $ f(r) =0$ admits complex conjugate roots\footnote{As a function of $r_e$, the disciminant interpolates between $\Delta(r_e=0)= -16G_N^2 M^2 \ell^8 (\ell^2 + 27 G_N^2 M^2$) to infinity. In particular, $\Delta \approx 256 \ell^6 r_e^6$ when $r_e \gg \ell $ and $\Delta \approx -16 G_N \ell^{10} M^2$ when $r_e \ll \ell$. In both cases, there is at least one pair of complex conjugate roots. }. We choose to parametrize them by $r_{1,2} = \hat{r}(1 \pm i  \epsilon)$ with $\epsilon>0$ and $ \hat{r}>0$. We can analytically solve for the other two roots, $r_3$ and $r_4$, and the parameter $\hat{r}$ by matching  the quadratic, cubic, quartic and constant contributions to $r^2 f(r)$. This parametrization is symmetric with respect to $\epsilon \mapsto -\epsilon$. Therefore, the expressions for $(r_3,r_4,\hat{r})$ will involve only even powers of $\epsilon$. In the near extremal limit ($\epsilon\ll 1$), we then approximate $f$ to order $\cO(\epsilon^4)$ by
\be \label{eq:approxf}
f(r)=\frac{1}{\ell^2} \left(\left(\frac{r-\hat{r}}{r}\right)^2+\left(\frac{\hat{r}\epsilon}{r}\right)^2\right)(r-r_3)(r-r_4)~,
\ee
with
\be
(r-r_3)(r-r_4) = \ell^2+r^2+2r\bar{r}+ 3 \bar{r}^2 - \frac{ \ell^2 (r+4\bar{r})  + 2 \bar{r}^2(r+6\bar{r})}{\ell^2 + 6\bar{r}^2}\bar{r}\epsilon^2 + \cO(\epsilon^4)~,
\ee
and
\be
\hat{r} = 
\bar{r}+\epsilon^2\frac{\bar{r}}{2\mathcal{C}(\bar{r})}\left(1+2\frac{\bar{r}^2}{\ell^2}\right)+\mathcal{O}(\epsilon^4)~.
\ee
Here $\mathcal{C}(r)$ is defined by
\be
\mathcal{C}(r) =6 \left(\frac{r}{\ell}\right)^2+1~.
\ee
In the region where $r-\bar{r} \ll \bar{r}$ and $\epsilon$ is small, we can approximate the metric (\ref{RNmetric}) as
\be
ds^2 = - \cC(\bar{r}) \left( \left(\frac{r-\bar{r}}{\bar{r}}\right)^2  + \epsilon^2 \right) d\tau^2 + \frac{dr^2}{ \cC(\bar{r}) \left( \left(\frac{r-\bar{r}}{\bar{r}}\right)^2  + \epsilon^2 \right)}+ \bar{r}^2 d \Omega_2^2~. \ 
\ee
By making the following identifications,
\eq{bhmatching}{
\rho = \frac{r - \bar{r}}{\epsilon \bar{r}}~~~{\rm and}~~~ t =  \mathcal{C}(\bar{r}) \frac{\tau \epsilon}{\bar{r}} ~,
}
the metric can be brought to global AdS$_2 \times \cS^2$ form 
\be
\label{eq:AdS2}
d s^2=\frac{\bar{r}^2}{\mathcal{C}(\bar{r})}\left(-(\rho^2+1)d t^2+\frac{d \rho^2}{\rho^2+1}\right)+\bar{r}^2 d\Omega_2^2~.
\ee
Following \cite{MMP}, we expect that in the wormhole region the solution is a slight perturbation of the near extremal RN black hole. We make the following gauge choice for our ansatz geometry in the throat
\be\label{eq:ansatzgeom}
d s^2=\frac{\bar{r}^2}{\mathcal{C}(\bar{r})}\left(-(1+\rho^2+\gamma)d t^2+\frac{d \rho^2}{1+\rho^2+\gamma}\right)+\bar{r}^2(1+\psi)d\Omega_2^2 ~,
\ee
where the functions $\psi(\rho)$ and $\gamma(\rho)$ are small fluctuations and $\bar{r}$ is given by (\ref{extremalcond2}). In these coordinates, the stress tensor contribution has the approximate form
\be\label{eq:quantumansatz}
\langle T^h_{\mu\nu}\rangle \approx-\frac{1}{2\pi^3}\frac{q\lambda(h)}{ \bar{r}^2}\text{diag}\left(1,\frac{1}{(1+\rho^2)^2},0,0\right)~.
\ee
The linearized Esintein's equations in this geometry are given by
\begin{align}
tt:&\quad\frac{\zeta}{1+\rho^2}+\psi(\rho)-\rho\psi^\prime(\rho)-(1+\rho^2)\psi^{\prime\prime}(\rho)=0~,\label{eq:tt}\\
\rho\rho:&\quad\frac{\zeta}{1+\rho^2}-\psi(\rho)+\rho\psi^\prime(\rho)=0~,\label{eq:rhorho}\\
\theta\theta:&\quad\frac{4}{\mathcal{C}(\bar{r})}\left(1+3\frac{\bar{r}^2}{\ell^2}\right)\psi(\rho)+\gamma^{\prime\prime}(\rho)+2\rho\psi^\prime(\rho)+(1+\rho^2)\psi^{\prime\prime}(\rho)=0~,\label{eq:thetatheta}\\
\phi\phi:&\quad \sin^2(\theta)\left(\frac{4}{\mathcal{C}(\bar{r})}\left(1+3\frac{\bar{r}^2}{\ell^2}\right)\psi(\rho)+\gamma^{\prime\prime}(\rho)+2\rho\psi^\prime(\rho)+(1+\rho^2)\psi^{\prime\prime}(\rho)\right)=0~,\label{eq:phiphi}
\end{align}
where $\zeta$ is a constant given by $\zeta=\frac{4G_N q\lambda(h)}{\pi^2\bar{r}^2}$\footnote{Note that we can write $q$ in terms of $\bar{r}$. This results in $\zeta=\frac{4 g\lambda(h)}{\pi^2\bar{r}}\sqrt{\frac{G_N\left(1+3\frac{\bar{r}^2}{\ell^2}\right)}{\pi}}$. From this we see that we can let $\zeta$ be small at finite $h$ and independent of the ratio between $\bar{r}$ and $\ell$.}. Note that the first two equations do not depend on $\gamma$. Therefore, we can find an expression for $\psi(\rho)$ by solving the first order equation (\ref{eq:rhorho}). This results in
\be\label{eq:psisol}
\psi(\rho)=\zeta(1+\rho\arctan(\rho))+c\rho~,
\ee
with $c$ an integration constant. A simple check shows that (\ref{eq:psisol}) also solves the $tt$ component of the Einstein equations (\ref{eq:tt}). By using the solution for $\psi(\rho)$, we can now use the angular components of the Einstein equations to solve for $\gamma(\rho)$. It turns out that 
\be\label{eq:gammasol}
\gamma(\rho)=-\frac{\zeta\left(1+4\frac{\bar{r}^2}{\ell^2}\right)}{\mathcal{C}(\bar{r})}\left(\rho^2+\rho(3+\rho^2)\arctan(\rho)-\log(1+\rho^2)\right)+c_1+\rho c_2~,
\ee
solves (\ref{eq:thetatheta}) and (\ref{eq:phiphi}). Integration constants can be set to zero by requiring that the geometry is invariant under $\rho\mapsto-\rho$ and by a redefinition of $\rho$ and $t$.

In the next subsection, we show that there is an overlapping region between the two solutions deep in the RN-AdS throat and construct the full wormhole geometry.
\subsection{Matching}

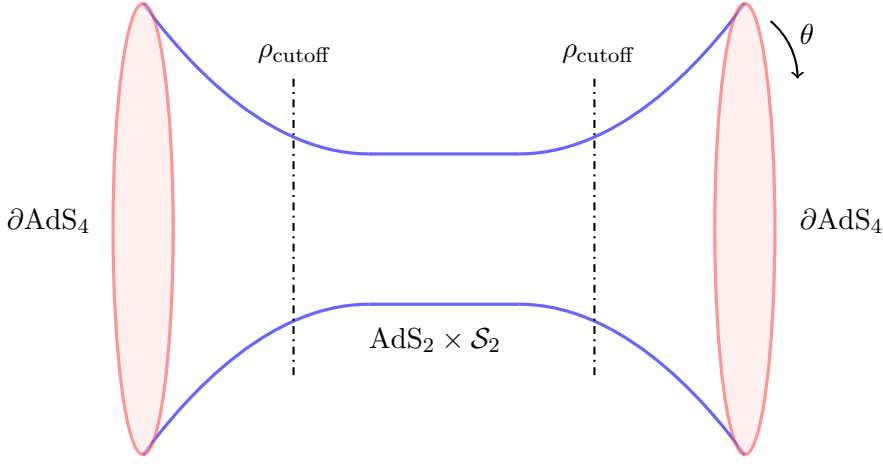
\begin{figure}
\centering
\begin{tikzpicture}
\draw[very thick, color=blue!60] (4,1) parabola (7,3.);
\draw[very thick, color=blue!60] (4,-1) parabola (7,-3);
\draw[very thick, color=blue!60] (2,1) -- (4,1);
\draw[very thick, color=blue!60] (4,-1) -- (2,-1);
\draw[very thick, color=blue!60] (2,1) parabola (-1,3);
\draw[very thick, color=blue!60] (2,-1) parabola (-1,-3);
\filldraw[color=red!70, fill=red!10, very thick, opacity=.6] (7,0) ellipse (.4cm and 3cm);
\filldraw[color=red!70, fill=red!10, very thick, opacity=.6] (-1,0) ellipse (.4cm and 3cm);
\draw[<-][thick](7.7,2) arc (0:50:1cm);
\draw [thick,dash dot] (5,2) -- (5,-2);
\draw [thick,dash dot] (1,2) -- (1,-2);
\begin{picture}(0,0)
\put(57,-45){${\rm{AdS}}_2 \times \cS_2$}
\put(-80,0){$\partial {\rm{AdS}}_4$}
\put(220,0){$\partial {\rm{AdS}}_4$}
\put(220,70){$\theta$}
\put(130,65){$\rho_{\rm cutoff}$}
\put(15,65){$\rho_{\rm cutoff}$}
\end{picture}
\end{tikzpicture}
\caption{Wormhole geometry: In the throat region the metric has the AdS$_2$  $\times {\mathcal{S}}^2$ form (\ref{eq:AdS2}) up to the cutoff located at $\rho\sim L/ \bar{r}$. Around this point, where the limits (\ref{eq:limits}) are satisfied, the geometry smoothly interpolates to near-extremal Reissner-Nordstr\"{o}m black holes in AdS$_4$. }
\label{Fig:wormhole}
\end{figure}

Intuitively, once the non-local coupling $h$ is turned on, the wormhole is formed and the throat acquires a certain finite length $L$ which we will determine below. Outside this range, the linearized solution found in the previous section will not be valid anymore. In fact, both perturbations $\psi$ and $\gamma$ increase with the value of $\rho$, as we approach the wormhole mouth. More precisely, we expect the slightly deformed solution to be valid up to values of $\rho$ for which the term $\zeta\rho^3$ is no longer subleading (since this is the leading order behaviour of $\gamma$). In the following, we consider $\rho$ to be large, but  $\zeta\rho$ small and fixed, and $\zeta$ small. We take the near-horizon limit of (\ref{eq:approxf}) 
\be\label{eq:fexpansion}
f(r)=\mathcal{C}(\bar{r})\epsilon^2+\mathcal{C}(\bar{r})\left(\frac{r-\bar{r}}{\bar{r}}\right)^2-\mathcal{C}\left(\bar{r}\right) \left( \frac{r-\bar{r}}{\bar{r}} \right) \epsilon^2-2\left(1+4\frac{\bar{r}^2}{\ell^2}\right)\left(\frac{r-\bar{r}}{\bar{r}}\right)^3+ \cdots~,
\ee
where we have expanded up to third order in $\epsilon$ and $\frac{r-\bar{r}}{\bar{r}}$ combined.
As a first approximation let us set
\be\label{eq:zerothordermatching}
\rho = \frac{L}{\bar{r}}\frac{r - \bar{r}}{\bar{r}},~~~~~ t = \mathcal{C}(\bar{r}) \frac{\tau}{L}~.
\ee
In the limit
\be\label{eq:limits}
 \rho \gg 1~, \quad \frac{L}{\bar{r}}\gg1~,\quad {\rm and} \quad \frac{r-\bar{r}}{\bar{r}}\ll1~,
\ee
equation (\ref{eq:zerothordermatching}) matches the order $\cO(2)$ of the unperturbed ansatz geometry. Here $L$ is an integration constant that denotes the rescaling between the $t$ and $\tau$ coordinates. Furthermore, by considering the relation between $\rho$ and $r$, one can see that $L$ is a measure up to which we can trust the ansatz; so that $\rho$ has a cutoff at $\rho\sim\frac{L}{\bar{r}}$. By comparing to the matching of the near-extremal Reissner-Nordstr\"{o}m black hole given in (\ref{eq:bhmatching}), we see that $L$ is connected to $\epsilon$ through\footnote{Recall that $\epsilon$ encodes how ``far" from extremality the near-extremal black hole metric is.}
\be\label{eq:Leps}
L=\frac{\bar{r}}{\epsilon}~.
\ee
By matching the angular coordinates we see that
\be\label{eq:psimatching}
r^2=\bar{r}^2(1+\psi(\rho))\quad\rightarrow\quad \frac{r-\bar{r}}{\bar{r}}=\frac{\psi(\rho)}{2}+\mathcal{O}\left(\psi^2\right)=\frac{\pi\zeta}{4}\rho+\mathcal{O}\left(\zeta^2\right)~,
\ee
where we have expanded $\sqrt{1+\psi(\rho)}$, and in the third equality we used the expansion of $\psi(\rho)$ at large $\rho$. Using (\ref{eq:zerothordermatching}) and (\ref{eq:psimatching}) we can find the value for $L$ by examining
\be\label{eq:L}
\rho d t=\mathcal{C}(\bar{r})\frac{r-\bar{r}}{\bar{r}^2}d\tau\quad\implies\quad L=\frac{d\tau}{dt}\mathcal{C}(\bar{r})=\rho\bar{r}\frac{\bar{r}}{r-\bar{r}}=\frac{4\bar{r}}{\pi\zeta}~.
\ee
One can easily see that with this value for $L$, (\ref{eq:zerothordermatching}) and (\ref{eq:psimatching}) are consistent with one another. This also gives a relation between the non-local coupling constant $h$ and $\epsilon$. With (\ref{eq:Leps}) and (\ref{eq:L}) we see that
\be\label{eq:heps}
\epsilon^2=\frac{\pi^2\zeta^2}{16}=
\frac{G_N g^2\lambda^2(h)}{\pi^3 \bar{r}^2} \left(1+3\frac{\bar{r}^2}{\ell^2}\right)=\frac{G_N^2q^2\lambda^2(h)}{\pi^2\bar{r}^4}~.
\ee
The matching of the time component of the geometry is discussed in Appendix \ref{sec:appmatching}. In this appendix we show that the equations above give a consistent matching between the Reissner-Nordstr\"{o}m geometry and the deformed $AdS_2\times \mathcal{S}^2$. A final comment we make concerning the matching is that the deformation $\gamma$ gives a correction to the range of the radial coordinate, which in turn leads to a correction to the stress tensor. However, this correction is of order $\zeta$, and therefore will not influence the matching\footnote{The correction can be calculated by considering $\Delta x=\int_{-\infty}^\infty \frac{dy}{g_{yy}}$, with $y$ the holographic coordinate,  resulting in $\Delta x=\pi \left(1+\zeta f(\bar{r})\right)$, for some function $f$. Since we consider $\zeta$ to be small, the matching is consistent. If we had taken this correction into account the stress tensor would have been given by $\langle T^h_{\mu\nu}\rangle=\frac{1}{\Delta x}\frac{q\lambda(h)}{ 2\pi^2 R^2}\text{diag}\left(1,1,0,0\right)$.}. The full wormhole geometry with the two regimes is schematically shown in Fig. \ref{Fig:wormhole}.

An important fact to notice about the wormhole solution we find is that there are three independent parameters: the charge, the non-local coupling and the AdS length by which the solution is determined. As soon as these three parameters are fixed, there is a unique, static and spherically symmetric wormhole geometry that solves Einstein's equations. At radii below the cutoff the geometry is that of deformed $\text{AdS}_2\times \cS^2$. As $\rho$ increases, the geometry smoothly interpolates to a near-extremal Reissner-Nordstr\"{o}m black hole in $\text{AdS}_4$. This black hole is characterized by its charge $r_e$, while its mass is given by 
\be \label{eq:whmass}
M_{WH}=M_{\text{ext}}+\Delta M,
\ee
with
\be\label{bhmass}
M_{\text{ext}}=\frac{\bar{r}}{G_N}+\frac{2\bar{r}^3}{G_N\ell^2},\quad\text{and}\quad\Delta M=-\frac{\bar{r}\epsilon^2}{2G_N}\mathcal{C}\left(\bar{r}\right)=
-\frac{g^2\lambda^2(h)}{2\pi^3\bar{r}}\left(1+3\frac{\bar{r}^2}{\ell^2}\right)\mathcal{C}(\bar{r})~,
\ee
where $M_{ext}$ is the mass of an extremal black hole.

Since it is not very pleasant to have a factor of $g$ in this formula, we use the definitions to rewrite this as 
\be
\Delta M =-\frac{G_Nq^2\lambda^2(h)}{2\pi^2\bar{r}^3}\mathcal{C}\left(\bar{r}\right) 
\sim -{q \lambda(h) \mathcal{C}\left(\bar{r}\right) \over \bar{r}}\zeta~.
\ee
Therefore, the black hole is indeed near-extremal, with mass just below the extremal mass. Coming from infinity, as an observer approaches the wormhole mouth, the observer would experience the geometry of a naked singularity. Of course, there is no actual singularity since as she gets closer to the center, the wormhole throat opens up and she traverses through the wormhole reaching the other side safely.


In the limit $\ell \gg r_e$, where the AdS radius is larger than the radii of the throats. The change in the mass due to the non-local coupling has the form
\be
\Delta M = -\frac{G_Nq^2 \lambda^2(h)}{2\pi^2r_e^3}~,
\ee
which has the same scaling as the binding energy, relative to the the energy of two disconnected extremal black holes, coming from the wormhole throat in the asymptotically flat case \cite{MMP}.

\subsection{Non-linear solution}
\begin{figure}[htbp]
\centering
\hspace*{3.4cm}\includegraphics[width=0.75\textwidth]{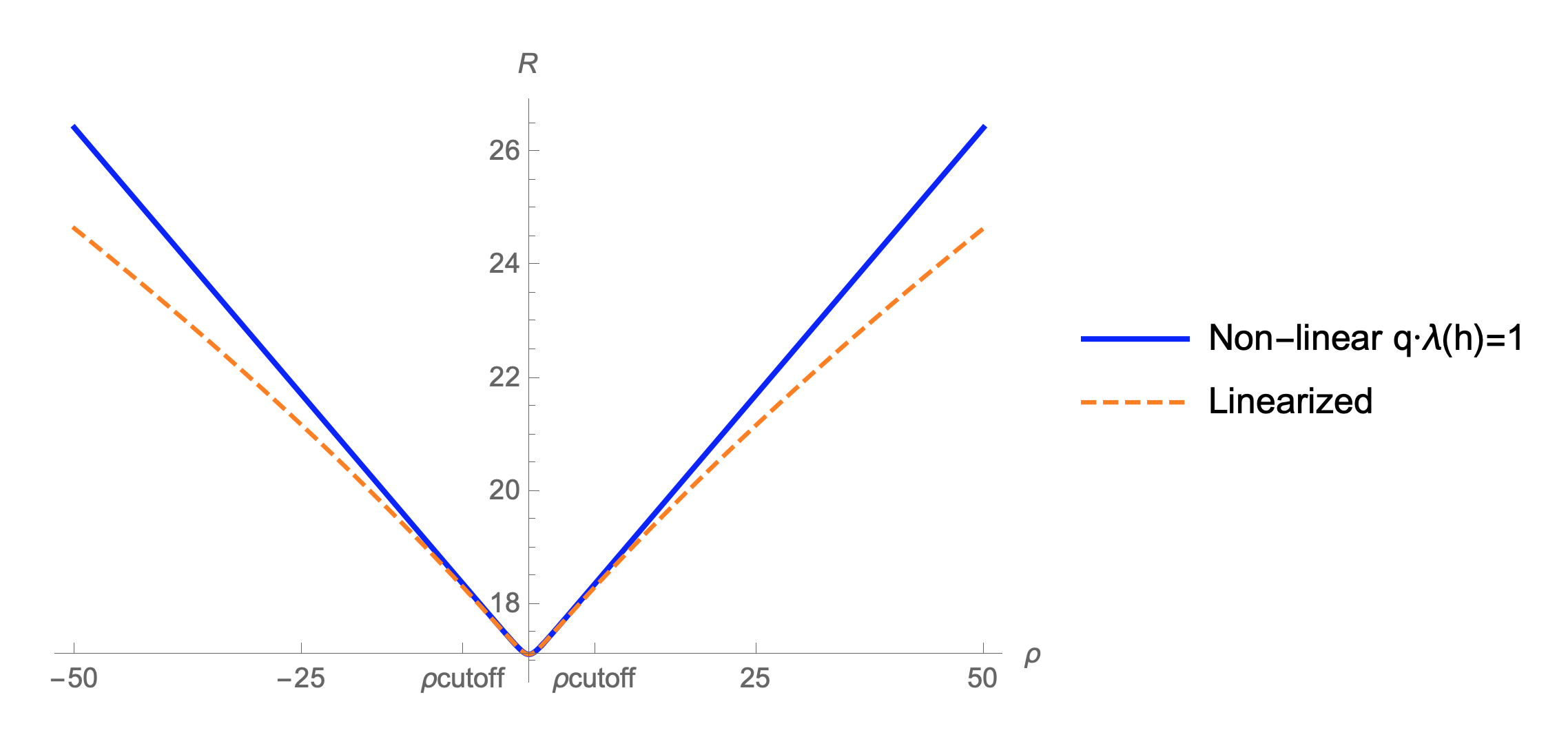}
\captionsetup{width=0.8\textwidth}
\caption{Solutions of $R(\rho)$ with parameters $\frac{G_N q^2}{g^2\ell^2}=0.01$, $\frac{G_Nq\lambda(h)}{\ell^2}=0.001$, and $\ell=100$. The initial condition is $R(0)=17$. For these parameters we expect agreement up to $\rho_\text{cutoff}= 7.2$. We see that for larger $\rho$ the linear solution starts to deviate.}
\label{fig:Rnumlinlarge}
\end{figure}

\begin{figure}[htbp]
\centering
\hspace*{3.4cm}\includegraphics[width=0.75\textwidth]{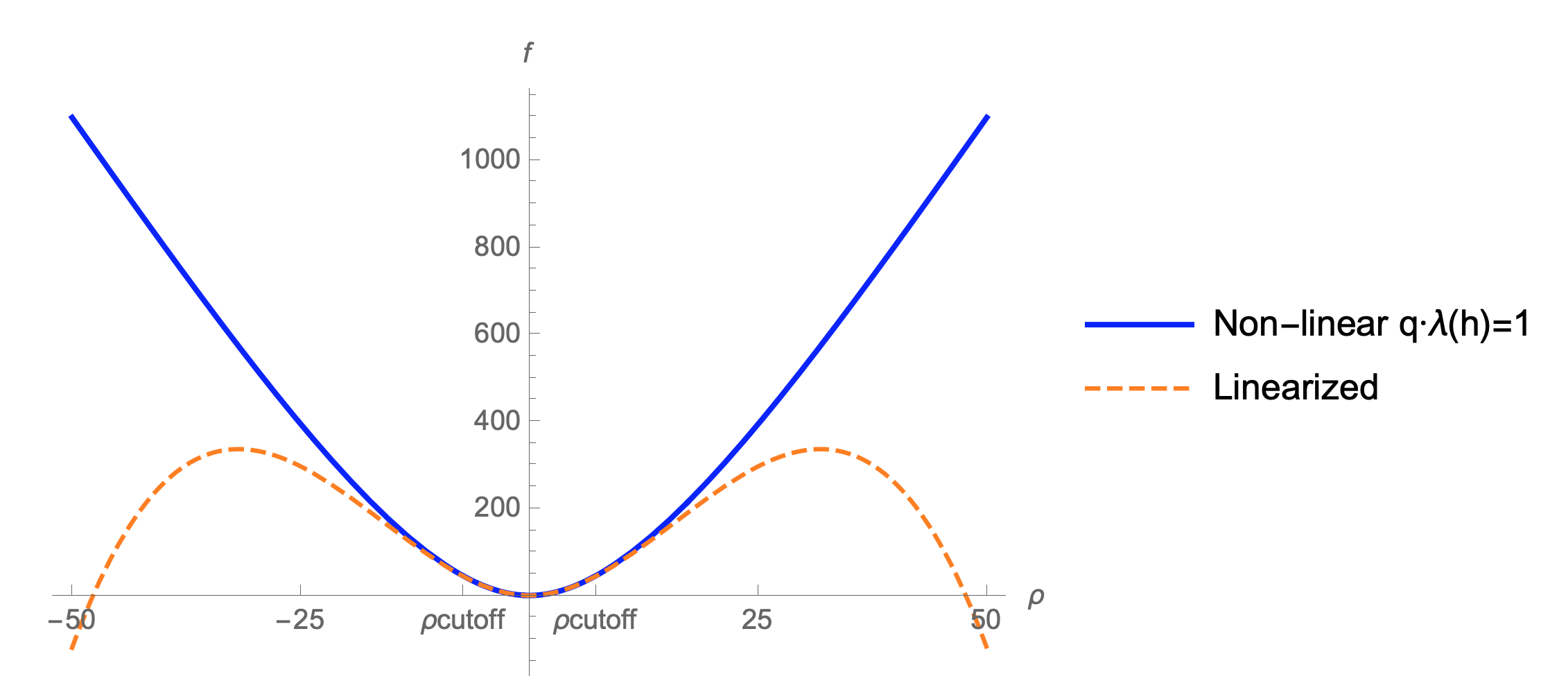}
\captionsetup{width=0.8\textwidth}
\caption{Solutions of $f(\rho)$ with parameters  $\frac{G_N q^2}{g^2\ell^2}=0.01$, $\frac{G_Nq\lambda(h)}{\ell^2}=0.001$, and $\ell=100$. The initial condition is $R(0)=17$. For these parameters we expect agreement up to $\rho_\text{cutoff}= 7.2$. We see that for larger $\rho$ the linear solution starts to deviate, and even becomes negative. Of course, in this region the RN AdS black hole dominates.}
\label{fig:fnumlinlarge}
\end{figure}

One might be concerned that the solution presented in the previous subsection only exists in the linearized analysis. We will proceed to find a similar solution to the full Einstein's equations. The geometry ansatz we will consider is the following
\be\label{numericansatz}
ds^2=\frac{\bar{r}^2}{\mathcal{C}(\bar{r})}\left(-f(\rho) dt^2+\frac{d\rho^2}{f(\rho)}\right)+R^2(\rho)d \Omega_2^2~,
\ee
$\rho \in [0,\pm \infty)$, $t \in (-\infty, \infty)$ and we have assumed the extremal value for the radius in the overall factor.  The non-zero components of the Einstein equations can be written as
\begin{align}
tt:\quad&\frac{3\bar{r}^2f(\rho)}{\mathcal{C}(\bar{r})\ell^2}-\frac{\pi G_Nq^2\bar{r}^2f(\rho)}{g^2\mathcal{C}(\bar{r})R^4(\rho)}+\frac{4G_Nq\lambda(h)}{\pi^2 R^2(\rho)}+\frac{\bar{r}^2f(\rho)}{\mathcal{C}(\bar{r})R^2(\rho)}\nonumber\\
&\quad-\frac{f(\rho)f^\prime(\rho)R^\prime(\rho)}{R(\rho)}-\frac{f^2(\rho)R^{\prime2}(\rho)}{R^2(\rho)}-\frac{2f^2(\rho)R^{\prime\prime}(\rho)}{R(\rho)}=0~,\label{eq:numericstt}\\
\rho\rho :\quad&-\frac{3\bar{r}^2}{\mathcal{C}(\bar{r})\ell^2f(\rho)}+\frac{\pi G_Nq^2\bar{r}^2}{g^2\mathcal{C}(\bar{r})f(\rho)R^4(\rho)}+\frac{4G_Nq\lambda(h)}{\pi^2 f^2(\rho)R^2(\rho)}\nonumber\\
&\quad-\frac{\bar{r}^2}{\mathcal{C}(\bar{r})f(\rho)R^2(\rho)}+\frac{f^\prime(\rho)R^\prime(\rho)}{f(\rho)R(\rho)}+\frac{R^{\prime2}(\rho)}{R^2(\rho)}=0~,\label{eq:numericsrhorho}\\
\theta\theta:\quad&-\frac{\pi G_N q^2}{g^2R^2(\rho)}-\frac{3R^2(\rho)}{\ell^2}+\frac{\mathcal{C}(\bar{r})R(\rho)f^\prime(\rho)R^\prime(\rho)}{\bar{r}^2}\nonumber\\
&\quad+\frac{\mathcal{C}(\bar{r})R^2(\rho)f^{\prime\prime}(\rho)}{2\bar{r}^2}+\frac{\mathcal{C}(\bar{r})f(\rho)R(\rho)R^{\prime\prime}(\rho)}{\bar{r}^2}=0~,\label{eq:numericsthth}\\
\phi\phi:\quad&\sin^2(\theta)\left(-\frac{\pi G_N q^2}{g^2R^2(\rho)}-\frac{3R^2(\rho)}{\ell^2}+\frac{\mathcal{C}(\bar{r})R(\rho)f^\prime(\rho)R^\prime(\rho)}{\bar{r}^2}\right)\nonumber\\
&\quad+\sin^2(\theta)\left(\frac{\mathcal{C}(\bar{r})R^2(\rho)f^{\prime\prime}(\rho)}{2\bar{r}^2}+\frac{\mathcal{C}(\bar{r})f(\rho)R(\rho)R^{\prime\prime}(\rho)}{\bar{r}^2}\right)=0~.\label{eq:numericsphiphi}
\end{align}
These differential equations depend on three independent physical parameters of the form
\be
\frac{G_N q^2}{g^2\ell^2},\quad\frac{G_N q\lambda(h)}{\ell^2},\quad\text{and}\quad\ell ~.
\ee
Since both functions $f$ and $R$ appear in the differential equations with two derivatives, there will be four integration constants. By requiring the solution to be symmetric around $\rho=0$ we fix two of those. Requiring this $\mathbb{Z}_2$ symmetry is equivalent to setting $f^\prime(0)=R^\prime(0)=0$. Furthermore we have the freedom to rescale the time coordinate. This allows us to pick $f(0)=1$. Now the constraint equation (\ref{eq:numericsrhorho}) fixes $R(0)$ in terms of $f(0)$. By these choices all integration constants are then fixed. Also note that due to spherical symmetry whenever the $\theta\theta$ equation is solved, the $\phi\phi$ equation is automatically satisfied. With the integration constants as mentioned, we can now solve the $tt$ and $\theta\theta$ equation numerically.  The results of solving the non-linear Einstein equations are shown in Figures \ref{fig:Rnumlinlarge} and \ref{fig:fnumlinlarge}. In order to compare with the linearized results, we pick the integration constants so that the non-linear and linear solutions agree at $\rho=0$. We should note however that the non-linear solution makes sense for other parameter values and integration constants as well. We expect the linear and non-linear results to agree up to $\vert \rho \lvert \sim\rho_{\text{cutoff}}= \frac{L}{\bar{r}}$. 
 As a final comment note that, as can be seen from Figure \ref{fig:Rnumlinlarge}, for large $\rho$ the numerical solution behaves as $R(\rho)\sim\left(\frac{L}{\bar{r}}\right)^{-1} \lvert \rho \lvert$, which is precisely what we expect in light of equation (\ref{eq:zerothordermatching}) and by the fact that away from the wormhole we expect the $\cS^2$ radius to be equal to $r$.

\FloatBarrier
\section{Thermodynamics}
This section contains a calculation of the on-shell Hamiltonian of the wormhole solution. We propose a Hamiltonian and show that the wormhole solution is the ground state for a region of parameter space. Furthermore we give a qualitative discussion of the thermodynamic stability of the wormhole solution in the (grand) canonical ensemble.

\subsection{Hamiltonian ground state}\label{sec:Hamiltonian}
We expect the wormhole geometry presented in the previous section to be dual to the asymptotic field theories in some particular entangled state. In particular, it should be dual to the ground state of a certain local Hamiltonian whose ground state is approximately the thermofield double state with chemical potential \cite{MQ, CFHL}.

From the gravity point of view, given the set of boundary conditions, Einstein's equations fill in the bulk geometry smoothly. We will consider three solutions with the same boundary conditions at zero temperature:  the wormhole, two disconnected black holes and empty AdS. Depending on the values of $\lambda(h)$ and $\mu$, there is a dominant saddle. For concreteness we will focus on the symmetric case where the total magnetic charge is $Q:=Q_R=-Q_L$ and the mass is $M:=M_R = M_L$. Of course, less symmetric cases can also be considered but we believe they will not dramatically modify the presented results. 

In a general covariant theory, the on-shell Hamiltonian can be computed as a boundary integral as follows 
\be\label{Hon-shell}
H[\zeta] = \int_{\partial \Sigma} d^2x \sqrt{\sigma} u^\alpha \zeta^\beta T_{\alpha \beta} ~,~~~ T_{\alpha \beta} := \frac{2}{\sqrt{- \gamma}} \frac{\delta S}{\delta \gamma^{\alpha \beta}}~,
\ee
where $T_{\alpha \beta}$ is the Brown-York stress tensor\footnote{In order to avoid IR divergences in AdS, we need to include counterterms in the purely gravitational part of the action (\ref{action1}). In d=3, they result in a modified stress tensor $T_{\alpha \beta}=K_{\alpha \beta} - K \gamma_{\alpha \beta} - \frac{2}{\ell} \gamma_{\alpha \beta} - \ell G_{\alpha \beta}$, where  $G_{\alpha \beta}$ is the Einstein tensor computed on the boundary induced metric $\gamma_{\alpha\beta}$ \cite{Balasubramanian:99}.}, $u^\alpha$ is the unit normal to a constant time hypersurface, $\zeta^\beta$ is the flow vector  and  $\sqrt{\sigma}$ is the volume element of the boundary at fixed time.
The energy of a gravitational solution is associated to time-translation symmetry, \ie, to the Killing vector $\zeta = \partial_\tau$. 

The wormhole solution presented in the last section is a solution to the action with interacting term 
\be\label{eq:actionint}
S_{\text{int}}= ih  \int d^3x \sqrt{-\gamma}  \left( \bar{\Psi}_-^R\Psi_+^L + \bar{\Psi}_{+}^{L} \Psi_-^R \right)~.
\ee
The interacting part of the boundary stress tensor then has the form
\be
T_{\alpha \beta}:= \frac{2}{\sqrt{- \gamma}} \frac{\delta S_{\text{int}}}{\delta \gamma^{\alpha \beta}} = ih \gamma_{\alpha \beta}    \left( \bar{\Psi}_-^R\Psi_+^L + \bar{\Psi}_{+}^{L} \Psi_-^R \right)~.
\ee
The metric close to the AdS$_4$ boundary at $r \rightarrow \infty$ and the time-like unit vector are of the form
\be\label{eq:asymptotic-metric}
d s^2=-f(r)d \tau^2+\frac{d r^2}{f(r)}+r^2 d \Omega_2^2 ~~~ \text{and} ~~~ u = \sqrt{f} d\tau~.
\ee
The $\mathbb{Z}_2$ symmetry along the radial direction allows us to define a notion of gravitational energy by applying the formula (\ref{Hon-shell}) at the asymptotic  AdS boundaries
\be
H_{\text{int}}:=H[\zeta^\tau] = -ih r^2 \sqrt{f} \int d^2\Omega \left( \bar{\Psi}_-^R\Psi_+^L + \bar{\Psi}_{+}^{L} \Psi_-^R \right)~.
\ee
At this point we need to evaluate the bulk spinors close to the asymptotic boundary. We can achieve this by evaluating the scaling factor close to the boundary\footnote{In the RN background (\ref{eq:asymptotic-metric}), far away from the horizon the metric is conformaly flat. The relation between the coordinates is $t=\frac{\tau}{L}\mathcal{C}(\bar{r})$, and $x = \int dr\frac{1}{L}\frac{1}{f(r)}\mathcal{C}(\bar{r})$, and the conformal factor equals $e^{2 \sigma} = \left(\frac{L}{\mathcal{C}(\bar{r})}\right)^2f$.}
\begin{align}
\bar{\Psi}_-^R\Psi_+^L & = \frac{e^{- \sigma}}{R^2} \left[ (\psi^R \otimes \eta)^\dagger \cP_- \gamma^1 \cP_+ \left(\psi^L \otimes \eta \right) \right] \\ & = \frac{\mathcal{C}(\bar{r})}{L r^2 \sqrt{f}} \left[ (\psi^R \otimes \eta)^\dagger \cP_- \gamma^1 \cP_+ \left(\psi^L \otimes \eta \right) \right]~,
\nonumber
\end{align}
and a similar expression for $\bar{\Psi}_{+}^{L} \Psi_-^R$.
We then find
\be
H_{\text{int}} = -ih \mathcal{C}(\bar{r})\frac{\epsilon}{\bar{r}}\int d^2\Omega  \left[ (\psi^R \otimes \eta)^\dagger \cP_- \gamma^1 \cP_+ \left(\psi^L \otimes \eta \right) + (\psi^L \otimes \eta)^\dagger \cP_+ \gamma^1 \cP_- \left(\psi^R \otimes \eta \right)   \right] ~.
\ee
We can compute the semi-classical interacting Hamiltonian in the state defined in (\ref{modes}) by computing the following expectation value
\be
\label{eq:Hintvev}
\langle H_{\text{int}} \rangle =  -ih \mathcal{C}(\bar{r})\frac{\epsilon}{\bar{r}} \int d^2\Omega  \left[ \langle (\psi^R \otimes \eta)^\dagger \cP_- \gamma^1 \cP_+ \left(\psi^L \otimes \eta \right) \rangle + \langle (\psi^L \otimes \eta)^\dagger \cP_+ \gamma^1 \cP_- \left(\psi^R \otimes \eta \right) \rangle   \right] ~.
\ee

We can evaluate the boundary integral by taking first the angular spinors on-shell. The correlators involved in 
(\ref{eq:Hintvev}) can be computed perturbatively in the limit where
$\lambda(h) \sim h$. They are explicitly given in Appendix \ref{sec:appcorrelators}. Finally, we obtain the result\footnote{Note that in the second equal sign we only take into account the terms up to order $h^2$, even though the expression in the middle contains a third order term. This is done to compare to the results of the previous section, which included terms up to second order.}
\be
\langle H_{\text{int}} \rangle = \mathcal{C}(\bar{r})\frac{\epsilon}{\bar{r}}\frac{2qh}{\pi} \left(-1 + \frac{4h}{\pi}\right)=4\Delta M+\mathcal{O}\left(h^3\right)~,
\ee
with $\Delta M$ given by (\ref{bhmass}).
In order to get the total wormhole energy, we need to add the energy asssociated to the non-interacting parts, \ie, of two near-extremal RN black holes
\be\label{eq:Hvev}
\langle H \rangle_{WH} = 2 M_{WH} + 4\Delta M- 2 \mu Q=2 M_{ext} + 6\Delta M- 2 \mu Q~,
\ee
where  $M_{ext}$ is the black hole extremal mass and $Q$ the extremal charge. In this equation, we have taken into account the change in energy due to the chemical potential $\mu$ for the asymptotic charges. This is similar to the electric case \cite{Hartnoll1}. 

Now that we understand the energy of the wormhole geometry, we would like to investigate whether it is the ground state of some Hamiltonian. 
We propose the following \emph{local boundary Hamiltonian}

\be\label{eq:hamiltonian}
H = H_L + H_R -\frac{ih}{\ell}  \int  d\Omega_2 \left(\mathbf{\bar{\Psi}}_-^R\mathbf{\Psi}_+^L + \mathbf{\bar{\Psi}}_+^L \mathbf{\Psi}_-^R  \right)+ \mu (Q_L-Q_R)~,
\ee
where $H_L$ and $H_R$ are the Hamiltonians associated to the boundary dual of the two identical original systems, and again we take into account the change in energy due to the chemical potential $\mu$ for the asymptotic charges. It is important to notice that equation (\ref{eq:hamiltonian}) is an expression written purely in terms of boundary data. In particular, we have defined the boundary spinors, denoted as $\mathbf{\Psi}$, by removing the scaling factor defined in (\ref{eq:extrapolate}), so that

\be
\Psi_{\pm}=R^{-\frac{3}{2}}\mathbf{\Psi}_{\pm}~.
\ee
Note that the interacting term is inspired by (\ref{eq:actionint}), which in terms of the boundary data can be written as

\be
S_{int}=\frac{ih}{\ell}  \int d\tau d\Omega_2 \left(\mathbf{\bar{\Psi}}_-^R\mathbf{\Psi}_+^L + \mathbf{\bar{\Psi}}_+^L \mathbf{\Psi}_-^R  \right).
\ee
The Hamiltonian determines the time-evolution with respect to the asymptotic time defined in (\ref{eq:asymptotic-metric}).
Note that the total charge of the field theories is conserved as a consequence of a global symmetry. 

Next, we consider the expectation value of the Hamiltonian for the different phases. First of all note that the expectation value of (\ref{eq:hamiltonian}) is precisely equal to (\ref{eq:Hvev}) for the wormhole solution, since that is the primary reason for the definition of (\ref{eq:hamiltonian}). Secondly, note that the empty AdS geometry has a vanishing Hamiltonian. Finally, we note that for the disconnected black holes, the interaction term of the Hamiltonian does not contribute to the energy. This can be seen from the fact that we can Wick rotate the RN black hole solution, after which the geometry is conformal to the disk. However, the conformal factor vanishes if the black hole is extremal. Since the correlators on the disk must be finite, the total contribution of the interacting part of the Hamiltonian must indeed be equal to zero.

\begin{figure}[htbp]
\centering
\hspace*{0cm}\includegraphics[width=0.6\textwidth]{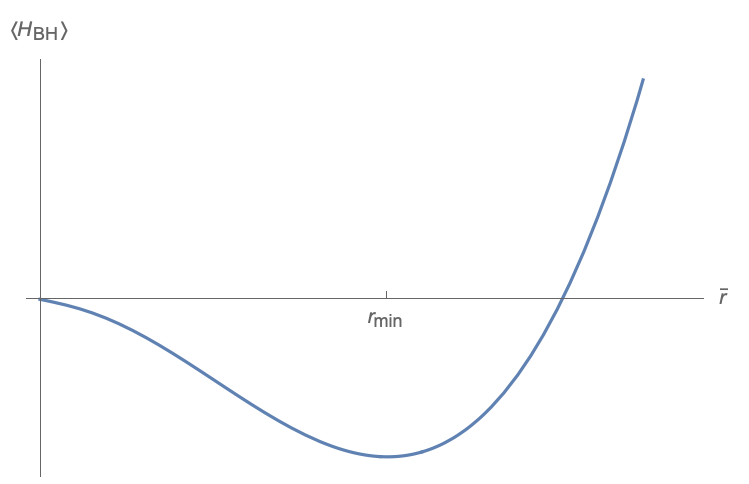}
\captionsetup{width=0.8\textwidth}
\caption{Expectation value of the Hamiltonian (\ref{eq:hamiltonian}) in the two-disconnected black holes phase for different values of $\bar{r}$. The minimum is located at $r_{\text{min}}$, which is given in (\ref{rmin}).}
\label{fig:Hamiltonian}
\end{figure}

\begin{figure}
\centering
\begin{tikzpicture}
\begin{picture}(0,0)
\fill[blue!10] (0,0) -- (0,0) --  (4,0) arc (1:90:4cm);
\fill[blue!10] (0,0) -- (4,0) -- (4,-3) -- (0,-3);
\draw[thick] (0,0) -- (4,0);
\draw[thick] (4,0) arc (1:90:4cm);
\draw[thick,->] (0,0) -- (0,5.5) node[anchor=south east] {$h$};
\draw[thick,->] (0,0) -- (7,0) node[anchor=north west] {$\mu$ };
\draw[thick,->] (0,0) -- (0,-3.2);
\draw[thick] (4,0) -- (4,-3);
\put(40,40){${\rm AdS}$}
\put(150,-40){${\rm BH}$}
\put(110,100){${\rm WH}$}
\put(120,-10){$\mu_c$}
\put(-15,110){$h_c$}
\put(-10,-5){$0$}
\end{picture}
\end{tikzpicture}
\caption{Diagram that shows the ground state of the Hamiltonian (\ref{eq:hamiltonian}) for different values of $h$ and $\mu$. Empty AdS is the dominant contribution at the origin up to the critical values 
$h_c=\frac{\bar{r}^2}{G_Nq}\sqrt{\frac{2\pi}{3\cC}\left(1+\frac{2\bar{r}^2}{\ell^2}\right)}$
and $\mu_c = m_p \sqrt{\pi}$ where the wormhole phase becomes the ground state. The point ($h=0$, $\mu=\mu_c$) is a triple point where the three phases meet. For negative values of $h$, there is a competition between the empty AdS and the black holes phases. Note that depending on the mass of the monopoles in the theory there could be a region in the diagram where the ground state is AdS with monopoles.}
\label{fig:phasediagram}
\end{figure}
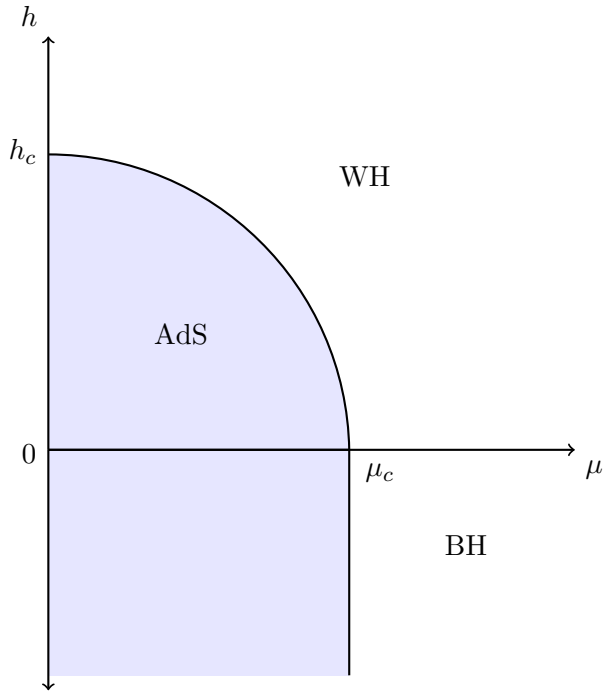

The difference between the wormhole and the extremal black holes phases is given by
\be
\langle H \rangle_{\text{2BH}} - \langle H \rangle_{\text{WH}} > 0~.
\ee
It is easy to see that $\langle H \rangle_{\text{2BH}}$ has a minimum at the point
\be\label{rmin}
\bar{r}=\frac{\ell}{\sqrt{3 \pi}} \sqrt{ \frac{\mu^2}{m_p^2}  -\pi }~,\quad\text{for}\quad \frac{\mu^2}{m_p^2} > \pi~,
\ee
for which $\langle H \rangle_{\text{2BH}} <\langle H \rangle_{\text{Vacuum}}=0$  (see Figure \ref{fig:Hamiltonian}). Then, 
\emph{the wormhole phase (where it exists) dominates the ground state for values of the chemical potential $\mu >\mu_c$}. The complete phase diagram is shown in Figure \ref{fig:phasediagram}. We see that the point $(h=0, \mu =\mu_c)$ is actually a \emph{triple point} where the three different phases meet. Intuitively, empty AdS is the dominant saddle for very small values of $h$, for which the wormhole has not been formed yet, and $\mu$ so that the charge contribution is negligible. For $h<0$ and $\mu>\mu_c$, the black holes phase is the dominant saddle. Alternatively, for positive values of the coupling and $\mu >\mu_c$, the wormhole phase will be the ground state of (\ref{eq:hamiltonian}).

\subsection{Stability}\label{sec:stability}

We briefly discuss possible instabilities of the solution. 
When considering scalar fields in a Reissner-N\"{o}rdstrom AdS background there are instabilities 
that lead to hairy black holes.
These instabilities can be understood, for near-extremal black holes, as originating from the difference between the Breitenl\"{o}hner Freedman bounds for $\text{AdS}_2$ and $\text{AdS}_4$; fields that are allowed tachyons in the asymptotic AdS$_4$ spacetime lead to instabilities in the AdS$_2$ near-horizon region.
Even though intuitively similar arguments would lead to fermionic instabilities, no evidence for the existence of fermionic hairy black holes has been found \cite{Dias_2020}. Since the argument crucially depends on the fact that there is an asymptotic $\text{AdS}_2$ geometry, we expect the same result to hold for the wormhole phase.

Besides investigating whether the wormhole solution is the ground state of the Hamiltonian (\ref{eq:hamiltonian}), one could also wonder whether it is the thermodynamically favored phase in one of the standard thermodynamic ensembles. We must couple the two CFT's in order  for a wormhole solution to exist; if signals can cross from one boundary to the other in the bulk, it must be possible to transfer information between the CFT's \cite{Ben1}.

Before coupling the CFT's, each CFT has a global $U(1)$ symmetry with an associated charge conservation. After coupling the theories, charge can flow from one to the other, so only a single $U(1)$ survives. In our conventions, the conserved charge us $Q_L + Q_R$. In our conventions, the wormhole solution has $Q_L = - Q_R$. One can picture magnetic field lines threading the wormhole, so this convention is natural. Therefore, the conserved charge for the wormhole is $Q_L + Q_R = 0$. In the standard construction of thermodynamic ensembles, one can only turn on a chemical potential for this conserved charge. The term $\mu (Q_L - Q_R) $ appearing in our Hamiltonian looks like a chemical potential, but it is not really, because $Q_L - Q_R$ is not conserved: it does not commute with the interaction term in the Hamiltonian.

One can ask whether the wormhole dominates one of the standard thermodynamic ensembles  at zero temperature\footnote{We should stress however, that our wormhole solution does exist for finite temperatures as well.}; for example, consider the canonical ensemble.
The different phases that should be compared are the following: empty AdS, two black holes, the wormhole solution.

The free energy  is 
\be\label{freeenergy}
F=E-TS~,
\ee
in the canonical ensemble. At zero temperature, the wormhole phase will have free energy equal to $F_{WH}=2M_{WH}$, the black hole phase has a free energy of twice the black hole mass, 
\be
F_{BH}=2 M_{BH} > F_{WH}~,  
\ee
while empty AdS has a vanishing free energy. Therefore, empty AdS dominates the ensemble. We leave the finite temperature discussion for future work. 

Finally, one could consider other ensembles in the hope of finding an ensemble in which the wormhole solution dominates. One candidate  is the grand canonical ensemble, with potential given by
\be\label{grandcan}
\Phi=E-TS-\mu Q~.
\ee
However, the wormhole phase has a conserved charge $Q=Q_L+Q_R=0$. Therefore, adding a term proportional to $Q$ will not change the potential of the wormhole phase. Because of this it can never dominate the ensemble. The only thing that we have achieved by changing ensembles is that there can be even more phases with a lower potential than the wormhole. Note that it is  not very clear how to interpret magnetic charges (and how to fix them) in the different ensembles. The electromagnetic duality suggests they should be treated in the same way as electric charges, but in the standard AdS/CFT context black holes with different electric charges are different states in the same theory, while black holes with different magnetic charges live in different theories. Presumably one can make a choice when imposing boundary conditions for the gauge field analogous to the standard vs alternate quantization for other light fields, and this choice determines whether electric or magnetic states live in the same theory. It would be interesting to understand this better, since our setup depends on the magnetic charges, but this subtlety is mostly orthogonal to our work here.

To summarize: the wormhole appears to be a stable solution that corresponds to the ground state of our Hamiltonian. However, it does not seem to arise as the dominant phase of one of the standard thermodynamic ensembles with chemical potential.

\FloatBarrier





\section{Discussion}
\label{S:Discussion}
We have found an eternal traversable wormhole in a four-dimensional AdS background. This geometry is a solution to the Einstein-Hilbert gravity action with negative cosmological constant, a $U(1)$ gauge field and massless fermions charged under the gauge field. 
To open up the wormhole we need negative energy, which we acquire by coupling the CFT's living on the two  boundaries of the spacetime. 

By calculating the backreaction of the negative energy on the geometry, we find a static traversable wormhole geometry with no horizons or singularities. 
This wormhole is dual to the ground state of a simple Hamiltonian for two coupled holographic CFT's. The parameters in the Hamiltonian are the chemical potential, the coupling strength, and the central charge. The wormhole dominates in some region of parameter space, while disconnected geometries dominate other regions.


Working in the semi-classical approximation, the authors of \cite{Ben1} proved that there are no traversable wormholes that preserve Poincar\'e invariance along the boundary field theory directions in more than two spacetime dimensions. The geometry found in this work evades this result because our solution is not  Poincar\'e invariant. 

There are a number of interesting future directions.
\paragraph{Traversable Wormholes in the lab; energy gap.} One possible application of our results is to build traversable wormholes in the lab by implementing our interacting Hamiltonian and allowing the system to cool to the ground state. For this process to be efficient, it is important the energy gap between the ground state and the first excited state is not too small. 

It would be interesting to carefully calculate the gap in this system. A rough estimate can be obtained by calculating the maximum redshift. Black holes have infinite redshift near the horizon and therefore support excitations with arbitrarily small energy in the semi-classical limit. 

Looking back at our `matching' section, we see that the black hole geometry is valid down to 
\begin{equation}
    r - \bar r \sim \epsilon L \sim \bar r~.
\end{equation}
At this location, the redshift is $f(r) \sim \mathcal{C}(\bar r)$. Inside this matching radius, the geometry is AdS$_2 \times S_2$. The relative redshift between the middle of the wormhole and the matching surface is
\begin{equation}
  {f_{\rm match} \over f_0} \sim \rho_{\rm match}^2 \sim {1 \over \epsilon^2}~.  
\end{equation}
Combining these results, and restoring units using the AdS radius, our guess is
\begin{equation}
    {\rm Gap} \sim {\mathcal{C}(\bar r)  \epsilon^2 \over \ell}~.
\end{equation}
This result looks concerningly small due to the $\epsilon^2$; however, $\mathcal{C}(r) = 1 + 6 \bar r^2/\ell^2$ is large for large black holes. We leave a fuller discussion and more reliable calculation for the future.

   \paragraph{RG flow.} Our bulk analysis is made convenient by the Weyl invariance of the massless fermions, which correspond to boundary operators of particular dimensions. If we think of the interaction term as an interaction in a single CFT, this term appears to be exactly marginal. It would be interesting to understand whether higher order corrections change the scaling dimension of this interaction, or more generally to understand the RG flow of our system.
  \paragraph{Supersymmetry.} Related to the RG flow, it would add a degree of theoretical control to realize the initial extremal black hole as a BPS state in a supersymmetric theory. Accomplishing this requires embedding our simple $U(1)$ theory in a theory with more conserved charges \cite{Romans_1992,Kunduri_2006,Gutowski_2004,Bhattacharyya_2011}.
 \paragraph{CFT state.} In the related construction of Cottrell et al. \cite{CFHL}, the state of the dual CFT was identified with the thermofield double state, while the bulk geometry was not under semiclassical control. In this paper, we have constructed a controlled traversable wormhole, but have not calculated the quantum state of the CFT in boundary variables. One may expect that it is a thermofield double type state, but note that the wormhole is also a solution at zero temperature, so it cannot be exactly the TFD. On the other hand, the bulk geometry clearly looks like a slightly superextremal Reissner Nordstrom black hole away from the wormhole mouth, giving a clear hint regarding the CFT state.
    \paragraph{Multi-mouth wormholes.}
    In the present work, we focused on asymptotically AdS$_4$ two-mouth traversable wormhole geometries. It might be interesting to extend our results and explore in the future the possibility of fourth dimensional multi-mouth wormholes similar to those studied in \cite{Emparan-Marolf, Balushi-Marolf}. In particular, the results of this paper might be used to understand explicitly the role played by multiparty entanglement in the wormhole's traversability. 
   \paragraph{\bf Information transfer.}
   Moreover, it would be interesting to investigate the amount of information that can be sent through this type of wormhole, in a similar fashion as in \cite{Maldacena:2017axo,Caceres:2018ehr,Freivogel:2019whb}.
\paragraph{Replica wormholes.} Finally, it has been found that two dimensional eternal traversable wormhole geometries contribute to the fine-grained entropy in the context of islands in de Sitter spacetime \cite{MaldaChen} (see also \cite{Hartman-Jiang, Mark2}). It would be interesting to understand whether more general set-ups in higher dimensions can be described with similar methods to those employed in this paper.

\section*{Acknowledgements}
We would like to thank Souvik Banerjee, Jan de Boer, Jackson R. Fliss, Victor Godet, Daniel Harlow, Jeremy van der Heijden, Diego Hofman, Daniel Jafferis, Bahman Najian, and Mark van Raamsdonk for discussions. 
BF, RE, and DN are supported by the ERC Consolidator Grant QUANTIVIOL. SB is supported by the European Research Council under the European
Unions Seventh Framework
Programme (FP7/2007-2013), ERC Grant agreement ADG 834878. This work
is part of the $\Delta$ITP consortium, a program of the NWO that is funded by the Dutch
Ministry of Education, Culture and Science (OCW). 
\section*{Appendices}
\addcontentsline{toc}{section}{Appendix}
\setcounter{subsection}{0}
\renewcommand{\thesubsection}{\Alph{subsection}}
\numberwithin{equation}{subsection}
\subsection{Propagators}\label{App:Propagators}
In this appendix we present the results for the propagators. Below, we show the derivation of \eqref{prop}
\begin{align}
\langle\psi_+^\dagger(x_-)\psi_+(x_-^\prime)\rangle&=\left\langle\sum_{k,j\in\mathbb{Z}}\frac{1}{\pi}\alpha_k^\dagger\alpha_je^{i\omega_jx_-^\prime-i\omega_kx_-}\right\rangle\nonumber\\
&=\left\langle\sum_{k,j\in\mathbb{Z}_{\ge0}}\frac{1}{\pi}\alpha_k^\dagger\alpha_je^{i\omega_jx_-^\prime-i\omega_kx_-}\right\rangle\nonumber\\
&=\left\langle\sum_{k,j\in\mathbb{Z}_{\ge0}}\frac{1}{\pi}(\delta_{k,j}-\alpha_j\alpha_k^\dagger)e^{i\omega_jx_-^\prime-i\omega_kx_-}\right\rangle\nonumber\\
&=\sum_{j\in\mathbb{Z}_{\ge0}}\frac{1}{\pi}e^{i\omega_j(x_-^\prime-x_-)}\nonumber\\
&=\sum_{j\in\mathbb{Z}_{\ge0}}\frac{1}{\pi}e^{i\left(\frac{2j+1}{2}\right)(x_-^\prime-x_-)}\left(1+(-1)^j\frac{2i\lambda(h)}{\pi}(x_-^\prime-x_-)\right)+\mathcal{O}\left((x_-^\prime-x_-)^2\right)\nonumber\\
&=\frac{1}{\pi}\frac{e^{\frac{i}{2}(x_-^\prime-x_-)}}{1-e^{i(x_-^\prime-x_-)}}+\frac{e^{\frac{i}{2}(x_-^\prime-x_-)}}{1+e^{i(x_-^\prime-x_-)}}\frac{2i\lambda(h)}{\pi^2}(x_-^\prime-x_-)+\mathcal{O}\left((x_-^\prime-x_-)^2\right)~.\label{eq:+*+prop}
\end{align}
The rest of the propagators can be derived in a similar fashion. We present the results
\be
\langle\psi_-^\dagger(x_+)\psi_-(x_+^\prime)\rangle=\frac{1}{\pi}\frac{e^{\frac{i}{2}(x_+^\prime-x_+)}}{1-e^{i(x_+^\prime-x_+)}}+\frac{e^{\frac{i}{2}(x_+^\prime-x_+)}}{1+e^{i(x_+^\prime-x_+)}}\frac{2i\lambda(h)}{\pi^2}(x_+^\prime-x_+)+\mathcal{O}\left((x_-^\prime-x_-)^2\right)~,
\ee

\be
\langle\psi_+^\dagger(x_-)\psi_-(x_+^\prime)\rangle=-\frac{1}{\pi}\frac{e^{\frac{i}{2}(x_+^\prime-x_-)}}{1+e^{i(x_+^\prime-x_-)}}-\frac{e^{\frac{i}{2}(x_+^\prime-x_-)}}{1-e^{i(x_+^\prime-x_-)}}\frac{2i\lambda(h)}{\pi^2}(x_+^\prime-x_-)+\mathcal{O}\left((x_-^\prime-x_-)^2\right)~,
\ee
and
\be
\langle\psi_-(x_+)\psi_+(x_-^\prime)\rangle=\langle\psi_-(x_+)\psi_-(x_+^\prime)\rangle=\langle\psi_+(x_-)\psi_+(x_-^\prime)\rangle=0~.
\ee


\subsection{Stress tensor}\label{sec:appT}

In this appendix we show the calculation of the stress tensor. First of all note that we can average over the angular directions by taking the spherical components on-shell in equation (\ref{stresstensorcomponents}). Since the spherical components are normalized such that $\int d^2 \Omega ~\bar{\eta}^{m} \eta^{n} = \delta_{m,n}$, averaging over the angular directions results in a factor of $\frac{1}{4\pi}$. Using this and point-splitting, the first line of (\ref{stresstensorcomponents}) becomes
\begin{align}
\langle T_{11}\rangle=\lim_{\substack{t^\prime\rightarrow t\\x^\prime\rightarrow x}}& \ \frac{1}{4\pi}\sum_{m,n}\int d^2\Omega\frac{ i}{2 R^2}\left(\partial_t^\prime-\partial_t\right)\left(\langle{\psi^m_+}^\dagger(x_-)\psi^n_+(x_-^\prime)\rangle+\langle\psi_-^{m\dagger}(x_+)\psi^n_-(x_+^\prime)\rangle\right)\eta^{m\dagger}\eta^n\nonumber \\
=\lim_{\substack{t^\prime\rightarrow t\\x^\prime\rightarrow x}}&\ \frac{qi}{8\pi R^2}\left(\partial_t^\prime-\partial_t\right)\left(\langle{\psi_+}^\dagger(x_-)\psi_+(x_-^\prime)\rangle+\langle\psi_-^{\dagger}(x_+)\psi_-(x_+^\prime)\rangle\right)~.
\end{align}
The renormalized $\langle T_{11}\rangle$ is found by subtracting the $h=0$ contribution from the $h\neq0$ expression as follows
\begin{align}
\langle T_{11}\rangle&=\langle T_{11}^{h\neq0}\rangle-\langle T_{11}^{h=0}\rangle\nonumber\\
&=\lim_{\substack{t^\prime\rightarrow t\\x^\prime\rightarrow x}}\frac{ qi}{8\pi R^2}\left(\partial_t^\prime-\partial_t\right)\left(\frac{e^{\frac{i}{2}(x_-^\prime-x_-)}}{1+e^{i(x_-^\prime-x_-)}}\frac{2i\lambda(h)}{\pi^2}(x_-^\prime-x_-)+\frac{e^{\frac{i}{2}(x_+^\prime-x_+)}}{1+e^{i(x_+^\prime-x_+)}}\frac{2i\lambda(h)}{\pi^2}(x_+^\prime-x_+)\right)\nonumber\\
&=\lim_{\substack{t^\prime\rightarrow t\\x^\prime\rightarrow x}}\frac{ qi}{8\pi R^2}\left(\frac{4i\lambda(h)}{\pi^2}+\frac{3i\lambda(h)}{2\pi^2}\left((t^\prime-t)^2+(x^\prime-x)^2\right)\right)\nonumber\\
&=-\frac{q\lambda(h)}{2 \pi^3 R^2}\label{eq:T11quantum}~,
\end{align}
 where we have omitted combined factors of $(t^\prime-t)$ and $(x^\prime-x)$ to higher orders. Similarly, we find that the rest of the renormalized components of the stress tensor are given by
\be
\langle T_{22}\rangle=-\frac{q\lambda(h)}{2 \pi^3 R^2},\quad\text{and}\quad\langle T_{33}\rangle=\langle T_{44}\rangle=0~.
\ee
It is easy to see that this contribution to the stress tensor is traceless
\eq{quantumTtraceless}{
\langle T_{\mu}^{\ \mu}\rangle=g^{\mu\nu}\langle T_{\mu\nu}\rangle=e^{-2\sigma}\left(\frac{q\lambda(h)}{2\pi^3 R^2}-\frac{q\lambda(h)}{2\pi^3 R^2}\right)=0~.
}
We can also check that the stress tensor is conserved. We will need the following components of $\nabla_\mu \langle T_{\rho \nu}\rangle$
\begin{align*}
\nabla_x\langle T_{xx}\rangle&=\partial_x\langle T_{xx}\rangle-\Gamma^\rho_{xx}\langle T_{\rho x}\rangle-\Gamma^\rho_{xx}\langle T_{x\rho}\rangle\\
&=\frac{q\lambda(h)R^\prime}{\pi^3 R^3}+\frac{q\lambda(h)\sigma^\prime}{\pi^3 R^2},\\
\nabla_t\langle T_{tx}\rangle&=\partial_t\langle T_{tx}\rangle-\Gamma^\rho_{tt}\langle T_{\rho x}\rangle-\Gamma^{\rho}_{tx}\langle T_{x\rho}\rangle\\
&=\frac{q\lambda(h)\sigma^\prime}{\pi^3 R^2},\\
\nabla_\theta \langle T_{\theta x}\rangle&=\partial_\theta\langle  T_{\theta x}\rangle-\Gamma^\rho_{\theta \theta }\langle T_{\rho x}\rangle-\Gamma^{\rho}_{\theta x}\langle T_{x\rho}\rangle\\
&=-\frac{q\lambda(h)R^\prime e^{-2\sigma}}{2\pi^3 R},\\
\nabla_\phi  \langle T_{\phi  x}\rangle&=\partial_\phi  \langle T_{\phi  x}\rangle-\Gamma^\rho_{\phi  \phi  }\langle T_{\rho x}\rangle-\Gamma^{\rho}_{\phi  x}\langle T_{x\rho}\rangle\\
&=-\frac{q\lambda(h)R^\prime e^{-2\sigma}\sin^2(\theta)}{2\pi^3 R}~,
\end{align*}
so that
\begin{align}
\nabla_\mu \langle T^{\mu}_{\ \nu}\rangle=&g^{\mu\rho}\nabla_\mu \langle T_{\rho \nu}\rangle\nonumber\\
=&g^{xx}\nabla_x\langle T_{xx}\rangle+g^{tt}\nabla_t\langle T_{tx}\rangle+g^{\theta\theta}\nabla_\theta \langle T_{\theta x}\rangle+g^{\phi\phi}\nabla_\phi \langle  T_{\phi  x}\rangle\nonumber\\
=&e^{-2\sigma}\left(\frac{q\lambda(h)R^\prime}{\pi^3 R^3}+\frac{q\lambda(h)h\sigma^\prime}{\pi^3 R^2}\right)\\
&-e^{-2\sigma}\frac{q\lambda(h)\sigma^\prime}{\pi^3 R^2}-\frac{1}{R^2}\frac{q\lambda(h)R^\prime e^{-2\sigma}}{2\pi^3 R}-\frac{1}{R^2\sin^2(\theta)}\frac{q\lambda(h)R^\prime e^{-2\sigma}\sin^2(\theta)}{2\pi^3 R}\nonumber\\
=&0.\label{quantumTconservation}
\end{align}
As a final check one can show that the $t,x$ component of the stress tensor is indeed equal to zero. First note that $T_{12}$ is equal to
\eq{T12_2}{
T_{12}= \frac{ i}{2 R^2}\left(\psi_+^\dagger\partial_-\psi_+-\psi_-^\dagger\partial_+\psi_--\partial_-\psi_+^\dagger\psi_++\partial_+\psi_-^\dagger\psi_-\right)\otimes\eta^\dagger\eta.
}
Using point-splitting this becomes
\begin{align}
\begin{split}
\langle T_{12}\rangle= &\lim_{\substack{t^\prime\rightarrow t\\x^\prime\rightarrow x}} \ \frac{1}{4\pi}\sum_{m,n}\int d^2\Omega 
\frac{ i}{2 R^2}\Big(\partial_-^\prime\langle\psi_+^{m\dagger}(x_-)\psi_+^n(x_-^\prime)\rangle-\partial_-\langle\psi_+^{m\dagger}(x_-)\psi_+^n(x_-^\prime)\rangle\\
& \quad \quad \quad \quad \quad\quad -\partial_+^\prime\langle\psi_-^{m\dagger}(x_+)\psi_-^n(x_+^\prime)\rangle-\partial_+\langle\psi_-^{m\dagger}(x_+)\psi_-^n(x_+^\prime)\rangle\Big)\eta^{m\dagger}\eta^n\end{split}\nonumber\\
\begin{split}
 =&\lim_{\substack{t^\prime\rightarrow t\\x^\prime\rightarrow x}} \ \frac{ qi}{8\pi R^2}\Big(\partial_-^\prime\langle\psi_+^\dagger(x_-)\psi_+(x_-^\prime)\rangle-\partial_-\langle\psi_+^\dagger(x_-)\psi_+(x_-^\prime)\rangle\\
& \quad \quad \quad \quad \quad\quad -\partial_+^\prime\langle\psi_-^\dagger(x_+)\psi_-(x_+^\prime)\rangle-\partial_+\langle\psi_-^\dagger(x_+)\psi_-(x_+^\prime)\rangle\Big)~,  
\end{split}
\end{align}
and we see that
\begin{align}
\langle T_{12}\rangle=&\langle T_{12}^{h\neq0}\rangle-\langle T_{12}^{h=0}\rangle\nonumber\\
\begin{split}
=&\lim_{\substack{t^\prime\rightarrow t\\x^\prime\rightarrow x}}\frac{ qi}{8\pi R^2}\Big(\partial_-^\prime\frac{e^{\frac{i}{2}(x_-^\prime-x_-)}}{1+e^{i(x_-^\prime-x_-)}}\frac{2\lambda(h)i}{\pi^2}(x_-^\prime-x_-)-\partial_-\frac{e^{\frac{i}{2}(x_-^\prime-x_-)}}{1+e^{i(x_-^\prime-x_-)}}\frac{2\lambda(h)i}{\pi^2}(x_-^\prime-x_-)\nonumber\\
&\quad  \quad-\partial_+^\prime\frac{e^{\frac{i}{2}(x_+^\prime-x_+)}}{1+e^{i(x_+^\prime-x_+)}}\frac{2\lambda(h)i}{\pi^2}(x_+^\prime-x_+)-\partial_+\frac{e^{\frac{i}{2}(x_+^\prime-x_+)}}{1+e^{i(x_+^\prime-x_+)}}\frac{2\lambda(h)i}{\pi^2}(x_+^\prime-x_+)\Big)\nonumber\\&\quad \quad +\mathcal{O}\left((x_-^\prime-x_-)^2\right)+\mathcal{O}\left((x_+^\prime-x_+)^2\right)\end{split}\nonumber\\
=&0~.
\end{align}

\subsection{Matching} \label{sec:appmatching}

Let us turn to the time component of the metrics. At large $\rho$ we can expand $\gamma(\rho)$ in the following way
\be\label{eq:largerho}
\gamma(\rho)=\frac{\zeta}{\mathcal{C}(\bar{r})}\left(1+4\frac{\bar{r}^2}{\ell^2}\right)\left(-\frac{\pi}{2}\rho^3-\frac{3\pi}{2}\rho+2\log(\rho)\right)+\cdots~.
\ee
In order to match the cubic term in $\rho$ to the $\left(\frac{r-\bar{r}}{\bar{r}}\right)^3$ term in (\ref{eq:fexpansion}) we will consider the following limit. We consider $\rho$ to be large, but $\zeta\rho$ small and fixed. In this limit $\psi$ is still given by (\ref{eq:psimatching}). Furthermore, from (\ref{eq:heps}) we see that we should consider $\epsilon$ to be of order $\zeta$. 
This limit corresponds to expanding $f$ at small $\frac{r-\bar{r}}{\bar{r}}$, but even smaller $\zeta\propto\epsilon$. More precisely, compared to (\ref{eq:fexpansion}) we still expand in $\frac{r-\bar{r}}{\bar{r}}$ up to third order. However, we only consider $\epsilon$ up to zeroth order 

\be\label{eq:fexpansion2}
f(r)=\mathcal{C}(\bar{r})\left(\frac{r-\bar{r}}{\bar{r}}\right)^2-2\left(1+4\frac{\bar{r}^2}{\ell^2}\right)\left(\frac{r-\bar{r}}{\bar{r}}\right)^3+\mathcal{O}\left(\left(\frac{r-\bar{r}}{\bar{r}}\right)^4\right)~.
\ee
We should now match the expansion (\ref{eq:fexpansion2}) to the following expression, where we will assume to be in the limit discussed above
\begin{align}
\frac{\bar{r}^2}{\mathcal{C}(\bar{r})}(1+\rho^2+\gamma(\rho))\frac{d t^2}{d\tau^2}&=\mathcal{C}(\bar{r})\left(\rho^2-\frac{\pi}{2}\rho^3\frac{\zeta}{\mathcal{C}(\bar{r})}\left(1+4\frac{\bar{r}^2}{\ell^2}\right)\right)\frac{1}{\rho^2}\left(\frac{r-\bar{r}}{\bar{r}}\right)^2\nonumber\\
&=\mathcal{C}(\bar{r})\left(\frac{r-\bar{r}}{\bar{r}}\right)^2-\frac{\pi}{2}\rho\left(\frac{r-\bar{r}}{\bar{r}}\right)^2\zeta\left(1+4\frac{\bar{r}^2}{\ell^2}\right)\nonumber\\
&=\mathcal{C}(\bar{r})\left(\frac{r-\bar{r}}{\bar{r}}\right)^2-2\left(1+4\frac{\bar{r}^2}{\ell^2}\right)\left(\frac{r-\bar{r}}{\bar{r}}\right)^3,\label{eq:gammamatching}
\end{align}
where in the third line we used (\ref{eq:psimatching}). We see that we recover (\ref{eq:fexpansion2}) up to third order.

\subsection{Correlators in 
$\langle H_{\text{int}} \rangle$}
\label{sec:appcorrelators}
In this appendix, we show explicitly the equal-time correlators involved in the computation of the semi-classical interacting Hamiltonian in terms of the $2D$ propagators presented in Appendix \ref{App:Propagators}. Note that in Appendix \ref{App:Propagators}, the propagators are expanded in $x-x'$, while here we need the location of the fields to approach opposing boundaries. However, the expressions obtained are still valid if we take $h$ small, and expand in $h$ instead. The equal-time correletors present in the Hamiltonian are given by\footnote{Note that we include higher orders than we need for the computation in the main text.}
\begin{align}
 \langle H_{\text{int}} \rangle  & \supset    -ih \cC(\bar{r}) \frac{\epsilon}{\bar{r}} \int d^2\Omega \langle \bar{\Psi}_-^R\Psi_+^L \rangle \nonumber\\   
    & =  -\frac{ih\epsilon}{2\bar{r}} \cC(\bar{r}) \int d^2\Omega \left( \langle - \psi_+^{R\dagger} \psi_+^L \rangle + i \langle  \psi_+^{R\dagger} \psi_-^L \rangle +i \langle \psi_-^{R \dagger} \psi_+^L \rangle + \langle \psi_-^{R\dagger} \psi_-^L \rangle\right) \eta_-^{\dagger}\eta_-\nonumber \\ 
   &  = \frac{qh\epsilon}{\pi\bar{r}} \cC(\bar{r})\left( -1 + \frac{4 h}{\pi} \right)+\mathcal{O}\left(h^4\right)~,
\end{align}
and
\begin{align}
 \langle H_{\text{int}} \rangle  & \supset   -ih \cC(\bar{r}) \frac{\epsilon}{\bar{r}} \int d^2\Omega \langle \bar{\Psi}_{+}^{L} \Psi_-^R \rangle\nonumber  \\ 
&= -\frac{ih\epsilon}{2\bar{r}} \cC(\bar{r})  \int d^2\Omega \left( \langle \psi_+^{L\dagger} \psi_+^R \rangle + i \langle  \psi_+^{L\dagger} \psi_-^R \rangle +i \langle \psi_-^{L \dagger} \psi_+^R \rangle - \langle \psi_-^{L\dagger} \psi_-^R \rangle\right) \eta_-^{\dagger}\eta_-\nonumber \\ 
&=\frac{qh\epsilon}{\pi\bar{r}} \cC(\bar{r}) \left( -1 + \frac{4 h}{\pi} \right)+\mathcal{O}\left(h^4\right)~.
\end{align}

\bibliographystyle{JHEP}
\bibliography{wormhole.bib}

\end{document}